\documentclass[
    reprint,
    amsmath,
    amssymb,
    aps,
    prd,
    superscriptaddress,
    floatfix,
    nofootinbib, 
    longbibliography
]{revtex4-2}

\usepackage{graphicx}
\usepackage{dcolumn}
\usepackage{bm}
\usepackage[utf8]{inputenc}
\usepackage[T1]{fontenc}
\usepackage{mathtools}

\usepackage{physics}

\usepackage{mathrsfs}
\usepackage{cancel}
\usepackage{siunitx}

\allowdisplaybreaks

\usepackage{booktabs}
\usepackage{tabularx}
\usepackage{comment}
\newcommand{\schwa}{\rotatebox[origin=c]{180}{e}}
\usepackage{microtype}

\usepackage{xcolor}
\definecolor{apslink}{RGB}{0,84,166}

\usepackage{hyperref}

\hypersetup{
    colorlinks   = true,
    linkcolor    = apslink,
    citecolor    = apslink,
    urlcolor     = apslink,
    breaklinks   = true,
    pdfstartview = {FitH}
}

\usepackage[capitalize]{cleveref}
\AtBeginDocument{\RenewCommandCopy\qty\SI}

\begin{document}

\title{Sensitivity of binary pulsar timing to spin-0 and spin-1 ultralight dark matter}

\author{Federico Huxhagen}
\email{fe.huxhagen@df.uba.ar}
\affiliation{Universidad de Buenos Aires, Facultad de Ciencias Exactas y Naturales, Departamento de Física, Buenos Aires, Argentina}
\affiliation{CONICET, Instituto de Física de Buenos Aires (IFIBA), Buenos Aires, Argentina}

\author{Diana L\'opez Nacir}
\email{dnacir@df.uba.ar}
\affiliation{Universidad de Buenos Aires, Facultad de Ciencias Exactas y Naturales, Departamento de Física, Buenos Aires, Argentina}
\affiliation{CONICET, Instituto de Física de Buenos Aires (IFIBA), Buenos Aires, Argentina}

\date{\today}

\begin{abstract}
If dark matter consists of ultralight bosons, on galactic scales it can be effectively described as a coherent classical field experiencing oscillations. Such a field could perturb the dynamics of celestial bodies via a direct coupling to ordinary matter, introducing signatures detectable through high-precision pulsar timing analysis. In this work, we extend a two-step Bayesian inference framework, originally developed for linearly coupled scalar ultralight dark matter (ULDM), to probe a quadratic scalar coupling and spin-1 vector dark matter. By explicitly marginalising over nuisance orbital parameters, our approach provides robust sensitivity limits that avoid the artificial overestimation often associated with direct fitting techniques. For quadratic scalar ULDM, we establish new constraints on the coupling $\beta$ for masses between $2 \times 10^{-22}$ eV and $2 \times 10^{-21}$ eV inaccessible to other experiments, while identifying mass regimes where the sensitivity is dominated by the orbital phase $\Psi'$ or the projected semi-major axis $x$. For vector ULDM, we characterize resonant signatures present even in circular orbits and obtain bounds on the coupling $g$ within the $10^{-23}$ eV to $10^{-18}$ eV range, yielding results within the same orders of magnitude as current laboratory and space-based experiments.
\end{abstract}

\maketitle

\section{Introduction}
\label{sec:intro}

Dark matter is one of the fundamental elements of the modern cosmological model and plays a central role in the formation and evolution of large-scale structures. Even though it comprises about 27\% of the energy content of the universe, its microscopic nature remains a mystery to this day \cite{Weinberg2008, darkmatter}. Among the different proposed scenarios, one alternative is that dark matter is composed of very light ($m <1$ eV) bosons, known as ultralight dark matter. Due to its macroscopic de Broglie wavelength, ULDM can effectively be described as a coherent classical field, oscillating at a frequency set by its mass across astrophysical scales \cite{MarshLibro}. 

In addition to the observed gravitational interaction, these dark matter fields may directly couple to ordinary matter, inducing perturbations that produce observable cumulative effects on highly stable astrophysical systems. In this context, binary pulsars are exceptional candidates for the study of ULDM. These rapidly rotating neutron stars act as incredibly precise clocks; particularly in binary systems, dynamical perturbations induced by external fields are directly reflected in the pulse time of arrival (TOA) measured on Earth. Various works have shown that gravitational and direct couplings leave characteristic imprints on pulsar timing data \cite{blas2019, Armaleo:2019txp, Blas2017Resonates, spin1, DeepNeuralULDM}, yielding competitive constraints on ULDM. 

However, most existing studies focus on individual systems and are restricted to the strictly resonant regime ---where the field frequency matches a multiple of the orbital frequency, leading to secular effects \cite{Rozner_2020}. While this regime provides strong constraints at discrete masses, it leaves large regions of parameter space unexplored and does not fully exploit the information available in multi-pulsar datasets. In view of upcoming radio facilities, such as the Square Kilometre Array \cite{SKA_Science_2015}, which are expected to monitor thousands of binary pulsars with unprecedented timing precision, it becomes essential to develop statistical frameworks capable of combining information from multiple systems.

In \cite{Kus2024}, a Bayesian method was developed to estimate the sensitivity of binary pulsars to ultralight dark matter beyond the strictly resonant regime, considering a spin-0 scalar field with a universal linear coupling to matter. This work extends that method to quadratically coupled spin-0 bosons as well as spin-1 vector ultralight dark matter. Such a quadratic coupling is well motivated in scenarios where the interaction respects the symmetry $\Phi \rightarrow -\Phi$, which forbids linear terms and makes $\Phi^2$ the leading-order contribution to the effective coupling. 

For the spin-1 case, existing constraints on the B-L coupling have been derived only within the strictly resonant regime \cite{spin1}, leaving the non-resonant parameter space largely unconstrained. The primary goal of this work is therefore to place constraints on the interaction between these fields and baryonic matter across a continuous range of dark matter masses.

In Sec. \ref{sec:theory}, the theoretical foundation of this work is presented. Sec. \ref{sec: timing} discusses the timing models used to map variations in the orbital dynamics to the measured times of arrival, whereas in Sec. \ref{sec: fuerzas perturbativas} the perturbing forces caused by ULDM are introduced and in Sec. \ref{sec: osculating orbits} their effect on the binary's orbital parameters is calculated. Then, Sec. \ref{sec:method} addresses the two steps which comprise the employed statistical method: Sec. \ref{sec: varianzas} details the variance estimation for the orbital parameters, while in Sec. \ref{sec: marginalizacion} we perform Bayesian inference, marginalising over the nuisance parameters. In Sec. \ref{sec:results} we present our main results, for both spin-0 (Sec. \ref{sec: results spin0}) and spin-1 (Sec. \ref{sec: results spin1}) ultralight dark matter. Our conclusions and outlook are discussed in Sec. \ref{sec:conclusions}. To make this article self-contained, we have also provided several appendices. Appendix \ref{ap: perturbed orbital parameters} introduces the osculating orbits equations as well as a Fourier expansion of the orbits in order to perform analytical calculations. Appendix \ref{ap: orbital perturb} contains the results of the integrals of the osculating equations. Appendix \ref{ap: varianzas} covers the variances obtained in the first step of the Bayesian method. In Appendix \ref{ap:marg}, the integrals required to perform the marginalisation are shown. Appendix \ref{ap: tablas} presents data on the observational precision considered. Appendix \ref{ap:one-step} outlines a comparison between our sensitivity estimates and those obtained through a one-step method. Lastly, Appendix \ref{ap: anomalia} develops a way to compute the sensitivity through the pulsar's true anomaly. Throughout this work, we adopt natural units where $\hbar = c = 1$.

\section{Orbital dynamics under ULDM perturbations}
\label{sec:theory}

\subsection{Timing models}
\label{sec: timing}

\begin{figure}[h]
    \centering
    \includegraphics[width=0.9\linewidth]{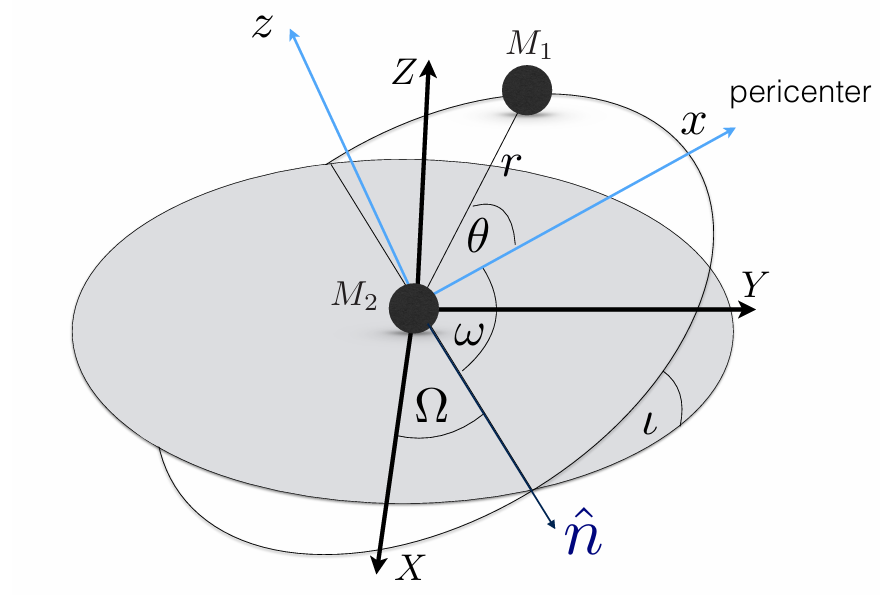}
    \caption{Description of Keplerian orbits in terms of the orbital elements, shown in the fundamental reference frame $(X,Y,Z)$. The Cartesian coordinate system $(x,y,z)$ and the polar coordinate system $(r,\theta,z)$ are also shown.}
    \label{fig: orbita kepleriana}
\end{figure}

To describe the orbital motion of the binary system, we can employ Kepler's laws, which allow the system to be parametrized in terms of six parameters. These are illustrated in Fig. \ref{fig: orbita kepleriana} and consist of: the semi-major axis $a$, the orbital eccentricity $e$, the longitude of periastron $\omega$, the orbital inclination $\iota$, the time of periastron passage $T_0$, and the longitude of the ascending node $\Omega$. The true anomaly $\theta$, which specifies the position of the pulsar along its orbit, is also indicated in the figure.

In this work, we adopt the Blandford-Teukolsky (BT) formalism \cite{BlandfordTeukolsky1976}, a standard framework for mapping binary orbital parameters to pulse arrival times while accounting for the main relativistic effects. 

Let $T_p$ be the proper time measured at the pulsar. The time of emission of the $N$-th pulse is related to the pulsar's rotation frequency $\nu$ by
\begin{equation}
     N = N_0 + \nu \, T_p + \dot{\nu} \frac{T_p^2}{2} + \ddot{\nu} \frac{T_p^3}{6},
     \label{eq: n-simo pulso}
\end{equation}
where $N_0$ is an arbitrary constant. The relation between the pulsar's proper time and the coordinate time $t$ is derived from the binary metric, accounting for gravitational redshift and transverse Doppler shift. Integrating $dT_p/dt$ yields
\begin{equation}
    T_p = t - \gamma_b \sin E,
    \label{eq: retardo de Einstein}
\end{equation}
where $\gamma_b = \frac{M_2^2(M_1+2M_2)}{a_1 \, (M_1+M_2)^2} \frac{P_b}{2 \pi} \, e$ is the Einstein delay. We also define the pulsar's semi-major axis as $a_1 = \frac{M_2}{M_1 + M_2} \, a$ and $E$ is the eccentric anomaly, which satisfies Kepler's equation 
\begin{equation}
E - e \sin E = \omega_b(t - T_0),
\end{equation} 
where $\omega_b = \frac{2\pi}{P_b}$ is the orbital frequency. 

While Eq. \eqref{eq: retardo de Einstein} relates the proper time of a hypothetical clock on the pulsar to the observer's coordinate time, connecting these results to terrestrial observations requires establishing the relationship between the pulse emission time $t_{\text{em}}$ and its arrival time $t_{\text{arr}}$. This temporal difference is primarily driven by geometric propagation, interstellar dispersion and relativistic delays.

In the BT model, the relationship between the infinite-frequency barycentric arrival time $t$ and the pulsar's proper time $T_p$ can be expressed as
\begin{equation}
    t = T_p + \alpha_b (\cos \tilde{E} - e) + (\beta_b + \gamma_b) \sin \tilde{E},
    \label{eq: relacion t y T_p}
\end{equation}
where $\alpha_b = x \sin \omega$ and $\beta_b = (1-e^2)^{1/2} x \cos \omega$, with $x=a_1 \, s$ being the projected semi-major axis and $s=\sin \iota$. In practice, $\beta_b$ and $\gamma_b$ are often combined into a single measurable parameter $\eta_b \equiv \beta_b + \gamma_b \approx \beta_b$. 

In order to write Eq. \eqref{eq: relacion t y T_p}, we have defined an auxiliary eccentric anomaly $\tilde{E}$ as
\begin{equation}
\tilde{E} - e \sin \tilde{E} = \frac{T_p}{\mathcal{P}} + \sigma.
\end{equation}

By perturbatively solving for the phase $N$ as a function of $t$ in Eq. \eqref{eq: n-simo pulso}, we obtain the BT timing model equation:
\begin{equation}
\begin{aligned}
 N &= N_0 + \nu t + \frac{1}{2}\dot{\nu} t^2 + \frac{1}{6}\ddot{\nu} t^3 - \nu f(E') - \dot{\nu} t f(E') \\
 &+ \frac{\nu \omega_b}{1 - e \cos E'} f(E') \left.\frac{df(z)}{dz}\right|_{z=E'},    
\end{aligned}
\label{eq: timing model BT}
\end{equation}
where $f(z) \equiv \alpha_b (\cos z - e) + \eta_b \sin z$. We can equivalently use two sets of orbital parameters: $\{a, e, \omega, s, T_0\}$ or $\{a, e, \alpha_b, \eta_b, \Theta'\}$. In addition to these, there are the parameters $N_0$, $\nu$, $\dot{\nu}$, and $\ddot{\nu}$, which are related to the pulsar's own rotation and are common to all timing models. Throughout this paper, we neglect the term corresponding to the Shapiro delay.

To reach Eq. \eqref{eq: timing model BT}, we have defined a new eccentric anomaly $E'$: 
\begin{equation}
E' - e \sin E' =  t / \mathcal{P} + \sigma = \int_{T_0}^t \omega_b \, dt' \equiv \Theta'.
\label{eq: definicion theta'}
\end{equation}

However, for systems with nearly circular orbits ($e \to 0$), the parameters $T_0$ and $\omega$ become ill-defined, since there is no periastron. To resolve this, we adopt the ELL1 model \cite{Lange2001}, which utilizes the Laplace-Lagrange parameters:
\begin{equation}
    \eta = e \sin \omega, \quad \kappa = e \cos \omega.
\end{equation}

In this framework, the time of periastron passage $T_0$ is replaced by the time of ascending node passage $T_{\text{asc}}$. We define the orbital phase $\Psi'$ in analogy to Eq. \eqref{eq: definicion theta'}: 
\begin{equation}
    \Psi'= \int_{T_{\text{asc}}}^t \omega_b \, dt' = E'-e \sin E' + \omega
    \label{eq: definicion Psi'}
\end{equation}
with $\omega = \int_{T_{\text{asc}}}^{T_0} \omega_b \, dt'$. Considering these new parameters, Eq. \eqref{eq: timing model BT} becomes
\begin{equation}
\begin{aligned}
N &= N_0 + \nu t + \frac{\dot{\nu}}{2}t^2 + \frac{\ddot{\nu}}{6} t^3 - \dot{\nu} \, x \, g(\Psi') \\ &+ \nu \omega_b \, \left. g(z) \frac{d g(z)}{dz}\right|_{z= \Psi'},
\end{aligned}
    \label{eq: ELL1 model}
\end{equation}
where $g(z) = -\frac{x \eta}{2} \cos(2 z) - \frac{3 x \eta}{2} + \frac{x \kappa}{2} \sin(2z) + x \sin z$. Eq. \eqref{eq: ELL1 model} defines the ELL1 timing model for circular orbits, with the set of parameters $\{ x, \eta, \kappa, s, \Psi' \}$ or with $T_{\text{asc}} \rightarrow \Psi'$.

\subsection{ULDM-induced perturbations}
\label{sec: fuerzas perturbativas}

As we have mentioned, the orbital dynamics of the binary system can be described with six orbital parameters: $a$, $e$, $\iota$, $\omega$, $\Omega$ and $T_0$. However, if an ultra-light dark matter field is present and directly coupled to ordinary matter, it perturbs the motion of the binary. 

The de Broglie wavelength of the ULDM field is set by the velocity dispersion of the halo $v_0$, which for the Milky Way we assume to be approximately $10^{-3}$ \cite{newfrontiers_lisanti}, so that
\begin{equation}
    \lambda_{\text{dB}} \sim 1.3 \times 10^{12} \, \text{km} \left( \frac{10^{-3}}{v_0} \right) \left( \frac{10^{-18} \, \text{eV}}{m} \right),
\end{equation}
which, in the mass range that we consider in this work (from $10^{-23}$ eV to $10^{-18}$ eV) is much greater than the typical size of the binary systems. Therefore, we can assume that the field oscillates coherently within each system, during a coherence time
\begin{equation}
    t_{\text{coh}} \sim 65 \, \text{y} \left( \frac{10^{-3}}{v_0} \right)^2 \left( \frac{10^{-18} \, \text{eV}}{m} \right),
\end{equation}
which is also considerably longer than the typical duration of observation campaigns, which can last up to a few decades. 

In the case of spin-0 dark matter, we effectively model it as a scalar field. We consider a quadratic, universal coupling to ordinary matter that modifies the masses of the binary components\footnote{A variation of the constituent masses also modifies the moment of inertia, and hence the spin frequency, of the neutron star, leaving an additional contribution in the residuals \cite{smarra2024constraints}. We restrict the present analysis to the orbital channel, so that our bounds are conservative with respect to the case in which both effects are included.}:
\begin{equation}
    \text{M}_i(\Phi) = \overline{\text{M}}_i \left(1 + \frac{\beta \, \Phi^2}{2}\right),
    \label{eq: masa cuadratica}
\end{equation}
where \(\overline{\text{M}}_i\) are the masses of the stars in the absence of the field ($i=1$ for the pulsar and $i=2$ for the companion), while $\beta$ is a constant which quantifies the strength of the coupling. The field can be expressed, on each binary system, as
\begin{equation}
    \Phi = \frac{\sqrt{2 \rho_{\text{DM}}}}{m} \, \varrho \, \cos (mt + \Upsilon),
\end{equation}
where $ \rho_{\text{DM}}\simeq 0.3 \frac{\text{GeV}}{\text{cm}^3}$ is the local dark matter density, $\Upsilon$ is a random phase and $\varrho$ is a random variable that comes from a Rayleigh distribution: 
\begin{equation}
    P(\varrho) = 2 \, \varrho \, e^{-\varrho^2}, 
\label{eq: rayleigh}
\end{equation}
which takes into account the stochasticity of the field. We will assume that the different binary systems of study are located in separate coherence patches, so that each system samples a different realization of $\varrho$ and $\Upsilon$.

The perturbed, non-relativistic equations of motion of the binary system, obtained through the Euler-Lagrange equations, are
\begin{subequations}
\begin{align}
    \ddot{\vec{r}} &= - G \, \text{M}_T \, \frac{\vec{r}}{r^3} - \beta \, \Phi \, \dot{\Phi} \,  \dot{\vec{r}}, \\
        \ddot{\vec{R}}&=- \dot{\vec{R}} \, \beta \, \Phi \, \dot{\Phi},
\end{align}
\end{subequations}
where $\text{M}_T = \text{M}_1 + \text{M}_2$ is the total mass of the system, $\vec{r}= \vec{r}_1- \vec{r}_2$ is the relative distance between the stars, $\vec{R}$ is the position of the center of mass and 
\begin{equation}
\vec{f}= - \beta \, \Phi \, \dot{\Phi} \,  \dot{\vec{r}}    
\end{equation}
is the perturbing acceleration induced by ULDM. 

As for the spin-1 scenario, we consider a vector field as defined in \cite{spin1}:
\begin{equation}
    \vec{A}(t) = -\frac{\sqrt{2\rho_{\text{DM}}}}{m} \, \varrho \, \cos(mt+\Upsilon)
    \begin{pmatrix}
        \sin\vartheta \cos\varphi \\
        \sin\vartheta \sin\varphi \\
        \cos\vartheta
    \end{pmatrix}
    \label{eq: vector field}
\end{equation}
where $\varrho$ and $\Upsilon$ are analogous to the scalar case and the spherical angles $\vartheta$ and $\varphi$ indicate the direction of the field. In this case, dark matter couples to the baryon number or B-L, therefore acting as a new composition-dependent fifth force which violates the equivalence principle. 

By introducing an interaction term 
\begin{equation}
\mathcal{L}_q = q_1 \, \vec{v}_1  \cdot \vec{A} +  q_2 \, \vec{v}_2  \cdot \vec{A},
    \label{eq: lagrangiano spin 1}
\end{equation}
where $q_i$ is the effective charge of each component of the binary and $\vec{v}_i$, the velocity, the Euler-Lagrange equations lead to a perturbing force per unit mass
\begin{equation}
    \vec{f} = -\mathcal{F}_0
    \begin{pmatrix}
        \sin\vartheta \cos(\varphi - \theta) \\
        \sin\vartheta \sin(\varphi - \theta) \\
        \cos\vartheta
    \end{pmatrix}
\end{equation}
where
\begin{equation}
\mathcal{F}_0 \equiv \frac{g \Delta c}{m_n} \sqrt{2 \rho_{\text{DM}}} \, \varrho \, \sin (m t + \Upsilon).    
\end{equation}
Here, $g$ is the coupling constant, $m_n$ is the neutron mass and $c_i= \frac{q_i \, m_n}{g \, \text{M}_i \, }$ is a phenomenological parameter that depends on the ratio between the baryonic and gravitational mass, as well as on the proton content of the star \cite{PRAKASH1997}. We denote the difference between the components of the system as $\Delta c = c_1 - c_2$. Since we aim to estimate an order of magnitude for the coupling constant $g$, we will take $\Delta c = 0.1$, as is customary in the literature \cite{delta_c}.

\subsection{Osculating orbits}
\label{sec: osculating orbits}

In what follows, we will adhere to the osculating orbits formalism. This means that we treat the ULDM effect as a perturbation, and we allow the orbital parameters to vary with time, following the equations derived through the variation of constants method in \cite{Gravity}. Using the expressions from Appendix \ref{ap: perturbed orbital parameters} for the time evolution of the orbital parameters, we find for the scalar case:
\begin{widetext}
\begin{subequations}
\label{eq: orbitas osculantes cuadratico}
\begin{align}
    \frac{\dot{a}}{a} &= -2 \beta \biggl\{ \frac{\omega_b \, a^2 \, \Phi^2 e \sin\theta}{2r^2\sqrt{1-e^2}} + \Phi \, \dot{\Phi} \, \frac{e^2 + 1 + 2e\cos\theta}{1-e^2} \biggr\}, \label{eq: a punto sobre a cuadratico completo} \\
    \dot{e} &= -\beta \biggl\{ \frac{\omega_b \, a^2 \, \sqrt{1-e^2} \, \Phi^2}{2 \, r^2} \, \sin \theta + 2 \, \Phi \, \dot{\Phi} \, (e + \cos \theta) \biggr\}, \\
     \dot{\Omega}& = \dot{\iota} = 0,  \\
    \dot{\varpi} &= \beta \left\{ \frac{\omega_b \, a^2 \, \sqrt{1-e^2} \, \Phi^2}{2 \, e \, r^2} \, \cos \theta - \frac{2 \, \Phi \, \dot{\Phi}}{e} \, \sin \theta \right\}, \\
    \dot{\epsilon}_1 &= \beta \biggl\{ \frac{\omega_b \, \Phi^2 \, (1 + e \cos \theta)}{1 - e^2} + (1 - \sqrt{1 - e^2}) \dot{\varpi} + 2 \, \Phi \, \dot{\Phi} \, e \sqrt{1 - e^2} \, \frac{\sin \theta}{1 - e^2} \, \frac{r}{a} \biggr\}, \label{eq: epsilon cuad}
\end{align}
\end{subequations}
\end{widetext}
where $\epsilon_1$ is defined as
\begin{equation}
\epsilon_1 \equiv \omega_b (t - T_0) + \omega + \Omega - \int_{T_0}^{t} dt'\, \omega_b \,,
\end{equation}
and it is related to $\Theta' = \Psi' - \omega$ through $\dot{\epsilon}_1 = \dot{\Omega} + \dot{\omega} + \dot{\Theta}' - \omega_b$.

Since $\Theta'$ has been shown to be the orbital parameter most sensitive to ULDM effects \cite{Kus2024}, we integrate $\dot{\Theta}'$ to find the accumulated phase shift. First, we expand the orbital frequency in the presence of ultralight dark matter as:
\begin{equation}
\begin{aligned}
    \omega_b^{\text{ULDM}} &= \sqrt{\frac{G \, \overline{\text{M}}_T \, (1 + \tfrac{1}{2} \beta \Phi^2)}{(a+ \delta a)^3}} \\
    &\simeq \omega_b  \left( 1 + \frac{1}{4} \beta \, \Phi^2 - \frac{3}{2} \frac{\delta a}{a}\right).
\end{aligned}
\label{eq: taylor wb}
\end{equation}

By using Eq. \eqref{eq: taylor wb} in Eq. \eqref{eq: epsilon cuad} and employing the Fourier series detailed in Appendix \ref{ap: perturbed orbital parameters}, we arrive at the final expression for the perturbation $\delta \Theta'$:
\begin{equation}
    \begin{aligned}
        \delta \Theta' = {} & \delta \Theta'_0 - \frac{3}{2} \omega_b \int_{T_0}^t \frac{\delta a}{a}(t') \, dt' + \frac{5}{8} \beta \, \omega_b \, \Phi_0^2 \, \tilde{t} \\
        & + \frac{5}{16} \, \frac{\beta \, \omega_b \, \Phi_0^2}{m} \biggl[ \sin(2 m t + 2 \Upsilon) \\
        & \hspace{65pt} - \sin(2 m T_0 + 2 \Upsilon) \biggr] \\
        & + \frac{\beta}{2} \sum_{n=1}^\infty \int_{T_0}^t \biggl[ A_n \, \Phi^2 \, \cos(n \omega_b \tilde{t}') \\
        & \hspace{55pt} + B_n \, \frac{d}{dt}(\Phi^2) \, \sin(n \omega_b \tilde{t'}) \biggr] dt'
    \end{aligned}
    \label{eq: cuadratico delta theta}
\end{equation}
where the coefficients are defined as 
\begin{subequations}
\begin{align}
    A_n &= 4 \, \omega_b \, J_n(ne) - 2\, \omega_b \frac{\sqrt{1-e^2}}{e} \, n \,  J_n'(ne), \\
    B_n &= \frac{4}{n} \, J_n(ne) + \frac{4}{e} \sqrt{1-e^2} \, J_n'(ne),
\end{align}
\end{subequations}
and we establish $\Phi_0 = \varrho \, \frac{\sqrt{2 \rho_{\text{DM}}}}{m}$ and $\tilde{t}=t-T_0$.

In Eq. \eqref{eq: cuadratico delta theta} we have included a correction $\delta \Theta '_0$, due to a potential error in the determination of the variable at $T_0$. In addition, we have allowed for a secular variation of the semi-major axis $a \rightarrow a + \dot{a}_0 \, (t - T_0)$, which can arise from different well studied effects such as the Shklovskii effect \cite{Shklovskii1970} ---due to the proper motion of the system--- or the acceleration of the binary due to the gravitational potential of the Milky Way \cite{DamourTaylor1991}. As well as that, we include a correction $\delta a_0$ analogous to $\delta \Theta'_0$. The term $\frac{\delta a}{a}(t')$, as well as the remaining variations of the orbital parameters, is provided in Appendix \ref{ap: orbital perturb}.

In the case of circular orbits, we can assume that the true anomaly evolves according to $\theta = \omega_b \, (t-T_0)$ (with $T_0 = T_{\text{asc}}+ \frac{\omega}{\omega_b}$) and that $r\rightarrow a$. Taking the limit $e \rightarrow 0$, Eq.\eqref{eq: orbitas osculantes cuadratico} becomes
\begin{subequations}
\begin{flalign}
    & \frac{\dot{a}}{a} = \frac{\dot{x}}{x} = -2 \beta \Phi \dot{\Phi} \label{eq: cuadratico circular semieje} && \\
    & \begin{aligned}
        \frac{\dot{\eta}}{\eta} = \frac{\dot{\kappa}}{\kappa} = \frac{\dot{e}}{e} = -\frac{\beta}{e} \biggl[ & \frac{\Phi^2}{2} \, \omega_b \, \sin(\omega_b \, \tilde{t}) \\ 
        &+ 2 \Phi \dot{\Phi} \, \cos(\omega_b \, \tilde{t}) \biggr]
    \end{aligned} && \\
    & \dot{\Omega} = \dot{\iota} = 0 && \\
    & \dot{\epsilon}_1 = \dot{\Psi}' - \omega_b = \beta \, \omega_b \, \Phi^2, \label{eq: cuadratico circular psi} &&
\end{flalign}    
\end{subequations}
In this case, we can directly integrate the osculating equations without the need of a Fourier decomposition, yielding: 
\begin{equation}
     \frac{\delta x}{x}
    = \frac{\delta x_{\mathrm{asc}} + \dot{x}_{\text{asc}}\,(t - T_{\mathrm{asc}})}{x}
    - \beta \big[ \Phi(t)^2 - \Phi(T_{\mathrm{asc}})^2 \big],
    \label{eq: resultado delta a sobre a}
\end{equation}
\begin{equation}
\begin{aligned}
\delta \Psi' &= \; \frac{11}{8} \beta \, \omega_b \, \Phi_0^2 \, (t- T_{\text{asc}}) \\
&- \frac{3}{2} \omega_b \frac{\delta a_{\text{asc}}}{a} (t- T_{\text{asc}}) - \frac{3}{4} \omega_b \frac{\dot{a}_{\text{asc}}}{a} (t- T_{\text{asc}})^2 \\
&+ \frac{11}{8} \frac{\beta \omega_b \Phi_0^2}{m} \cos\left[ m(t + T_{\text{asc}}) + 2\Upsilon \right] \sin(m \tilde{t}) \\
&- \frac{3}{2} \, \omega_b \, \beta \, \Phi^2(T_{\text{asc}}) (t- T_{\text{asc}}) + \delta \Psi ' _{\text{asc}}\,
\end{aligned}
\label{eq: cuadratico variacion psi}
\end{equation}
for the projected semi-major axis and $\Psi'$ respectively. 

In the case of a vector field, the interaction depends on the relative orientation between the pulsar's position and the dark matter field direction. The derivatives of the orbital elements are obtained in a similar way to the scalar case:
\par\medskip\begin{widetext}\begin{subequations}
\label{eq: orbitas osculantes spin 1}
\begin{align}
    \dot{a} &= -\frac{2 \mathcal{F}_0}{\omega_b} \sin\vartheta \biggl[ \frac{e \sin\theta \cos(\varphi-\theta)}{\sqrt{1-e^2}} + \frac{a\sqrt{1-e^2}}{r} \sin(\varphi-\theta) \biggr], \label{eq: adot exacto spin 1} \\
    \dot{\Omega} &= - \frac{\mathcal{F}_0 \,\cos\vartheta}{a\,\omega_b}\,\frac{\sin\theta \cos\omega + \cos\theta \sin\omega}{\sin\iota \,\sqrt{1-e^2}}\, \frac{r}{a}, \\
    \dot{\iota} \, &= -\frac{\mathcal{F}_0\cos\vartheta}{a\,\omega_b}\,
\frac{\cos(\theta+\omega)}{\sqrt{1-e^2}}\,\frac{r}{a}, \\
    \dot{\varpi} &= - \frac{\sqrt{1-e^2}\mathcal{F}_0}{a \, e \, \omega_b} \sin\vartheta \biggl[ -\cos\theta \cos(\varphi-\theta) + \sin\theta \sin(\varphi-\theta) \left(1 + \frac{r/a}{1-e^2}\right) \biggr] + 2\sin^2\left(\tfrac{\iota}{2}\right)\dot{\Omega}, \\
    \dot{\epsilon}_1 &= \frac{2 \mathcal{F}_0}{a \omega_b} \frac{r}{a} \sin\vartheta \cos(\theta-\varphi) + \big[1-\sqrt{1-e^2}\big]\dot{\varpi} + 2\sqrt{1-e^2}\sin^2\left(\tfrac{\iota}{2}\right)\dot{\Omega}.
\end{align}
\end{subequations}
\end{widetext}
\par\medskip
 
Following the same logic as before, we describe the accumulated phase shift through the parameter $\Theta'$, defined as:
\begin{equation}
\begin{aligned}
    \delta \Theta' &= \delta \Theta'_0 - \frac{3}{2}\,\omega_b \int_{T_0}^t \frac{\delta a}{a}(t')\,dt' 
    + \int_{T_0}^t \dot{\epsilon}_1(t')\,dt' \\ &- \int_{T_0}^t \dot{\varpi}(t')\,dt' \,,
    \label{eq: delta theta' spin 1}
\end{aligned}
\end{equation}
where the first integral accounts for the ULDM-induced variation in the orbital frequency, while the others represent direct perturbations to the osculating elements. 

In the circular orbit limit ($e \to 0$, $r \to a$), the semi-major axis variation $\delta a(t)$ admits a closed-form analytical expression:
\begin{flalign}
    & \begin{aligned}
        \delta a(t) = & \frac{g\,\Delta c}{\omega_b \, m_n}\,\sqrt{2\rho_{\mathrm{DM}}}\,\varrho\,\sin\vartheta \\
        &  \left[C^{(a)} \sin \Upsilon_{\mathrm{asc}} + S^{(a)} \cos \Upsilon_{\mathrm{asc}} \right],
    \end{aligned} && \label{eq: deltaa_final}
\end{flalign}
where the functions $S^{(a)}$ and $C^{(a)}$ are given by
\begin{subequations}
\begin{align}
    S^{(a)} &= \sum_{\pm} \mp \frac{\sin (m_{\pm} \tilde{t} \mp \varphi_{\mathrm{asc}}) \pm \sin \varphi_{\mathrm{asc}}}{m_{\pm}} \\
    C^{(a)} &= \sum_{\pm} \mp \frac{\cos (m_{\pm} \tilde{t} \mp \varphi_{\mathrm{asc}}) - \cos \varphi_{\mathrm{asc}}}{m_{\pm}},
\end{align}
\label{eq: spectral_functions}
\end{subequations}
\noindent \hspace{-0.2cm} with $m_{\pm} \equiv m \pm \omega_b$, $\Upsilon_{\text{asc}}= m \, T_{\text{asc}}+\Upsilon$ and $\varphi_{\text{asc}}= \varphi + \omega$. The denominators in Eq. \eqref{eq: spectral_functions} explicitly show the resonant nature of the interaction: as the ULDM mass $m$ approaches the orbital frequency $\omega_b$, the functions $S^{(a)}$ and $C^{(a)}$ develop a secular growth. This confirms that binary systems can impose restrictive constraints even in the circular limit due to these resonant effects. 

The corresponding phase perturbation $\delta \Psi'$ follows the structure of Eq. \eqref{eq: delta theta' spin 1}:
\begin{samepage}
\begin{align}
\delta \Psi'(t)
    &= \delta \Psi'_0
    -\frac{3}{2}\,\omega_b \notag
    \int_{T_{\text{asc}}}^t \frac{\delta a}{a}(t')\,dt'
    + \int_{T_{\text{asc}}}^t \dot{\epsilon}_1(t')\,dt'
    \\ &- \int_{T_{\text{asc}}}^t \dot{\Omega}(t')\,dt'. 
        \label{eq: delta_Psi_general_spin1}
\end{align}
\end{samepage}

Considering only the effect of ULDM on the orbital frequency accurately captures the on-resonance sensitivity and the overall order of magnitude, as was studied in \cite{Kus2024}. However, numerical integration is required to compute the full signal from all terms in Eq. \eqref{eq: delta_Psi_general_spin1}.

\section{Statistical methodology}
\label{sec:method}

In this work, we employ a two-step inference approach to analyze the effects of ULDM on pulsar timing, as initially introduced in \cite{Kus2024} and following the procedure defined in \cite{Moore2015}. This approach is motivated by the inherently stochastic nature of the ULDM field: the local amplitude $\Phi_0$ follows a Rayleigh distribution, while the phase $\Upsilon$ is uniformly distributed in $[0, 2\pi)$. Since these parameters cannot be determined a priori, they are treated as nuisance parameters alongside other potential effects and noise processes that may be present in the timing signal. Our objective is to quantify the sensitivity to the coupling constant $\beta$ (or $g$) independently of the specific field realization. We therefore adopt a Bayesian framework, performing a marginalisation over these stochastic unknowns and additional signal components.

\subsection{Variance estimation}
\label{sec: varianzas}
Suppose that we have an initial approximation $N_0^{(1)}, \nu^{(1)}, \dot{\nu}^{(1)}, \ldots$ for the orbital parameters of the system. This can be obtained in various ways, such as fitting pulse arrival time data using tools such as \texttt{TEMPO2} \cite{Hobbs2006} or \texttt{PINT} \cite{Luo_2021}. These are widely used software packages in pulsar timing analysis that allow the fitting of timing model parameters to observational data with high precision.

This defines the \textit{calculated} values as
\begin{equation}
N^{(1)} = N\!\left(t; N_0^{(1)}, \nu^{(1)}, \cdots \right),
\label{eq: 2.42}
\end{equation}
while the \textit{observed} values are
\begin{equation}
N = N(t; N_0, \nu, \ldots),
\label{eq: 2.43}
\end{equation}
and they depend on the true values of the parameters. The residuals $R(t)$ (in units of time) are defined by
\begin{equation}
\begin{aligned}
- \nu^{(1)} R(t) \equiv N - N^{(1)} 
&= \left. \frac{\partial N}{\partial N_0} \right|_1 \delta N_0 + \left. \frac{\partial N}{\partial \nu} \right|_1 \delta \nu \\ 
& \quad + \cdots + \left. \frac{\partial N}{\partial \gamma_b} \right|_1 \delta \gamma_b ,
\end{aligned}
\label{eq: 2.44-45}
\end{equation}
where $\delta N_0, \delta \nu, \ldots, \delta \gamma_b$ are estimates of the parameter corrections. 

Explicitly, the residual function $R(t)$ that relates the observed times of arrival to the prediction according to Blandford-Teukolsky is 
\begin{equation}
\begin{aligned}
R(t) = \, &\delta K - \frac{\partial N}{\partial a } \, \frac{\delta a }{\nu} - \frac{\partial N}{\partial e } \, \frac{\delta e }{\nu} - \frac{\partial N}{\partial \alpha_b } \, \frac{\delta \alpha_b }{\nu} \\&- \frac{\partial N}{\partial \eta_b} \frac{\delta \eta_b}{\nu} - \frac{\partial N}{\partial \Theta'} \frac{\delta \Theta'}{\nu},
\end{aligned}
    \label{eq: modelo BT}
\end{equation}
while in the case of circular orbits (ELL1 model), 
\begin{equation}
\begin{aligned}
R(t) = &\delta K - \frac{\partial N}{\partial x } \, \frac{\delta x }{\nu} - \frac{\partial N}{\partial \eta } \, \frac{\delta \eta}{\nu} - \frac{\partial N}{\partial \kappa } \, \frac{\delta \kappa }{\nu} \\&- \frac{\partial N}{\partial s} \frac{\delta s}{\nu} - \frac{\partial N}{\partial \Psi'} \frac{\delta \Psi'}{\nu}.
\end{aligned}
    \label{eq: modelo ELL1}
\end{equation}

To simplify notation, we define a vector $\mathbf{\delta S}$ containing the model parameter residuals that can be measured in both the BT and ELL1 models. Specifically, we have 
\[\mathbf{\delta S}_{\mathrm{BT}} \equiv \{\delta K, \, \delta a, \, \delta e, \, \delta \alpha_b, \, \delta \eta_b, \, \delta \Theta' \},\] 
\[ \mathbf{\delta S}_{\mathrm{ELL1}} \equiv \{\delta K, \, \delta s, \, \delta x, \, \delta \eta, \, \delta \kappa, \, \delta \Psi' \}.\] 

We now assume that observations are performed at regular time intervals, with a cadence (number of measurements per time interval) $\dot{n}$ over a duration $d$, under the condition $P_b \ll d \ll T_{\mathrm{obs}}$, where $P_b$ is the orbital period and $T_{\mathrm{obs}}$ the total observation time. Here $d$ is the length of the sub-intervals into which $T_{\mathrm{obs}}$ is partitioned for our method (each containing $\dot n d$ TOAs).\footnote{Do not confuse $d$ with the per-session integration time used to build a single pulse profile.} The condition $d \ll T_{\mathrm{obs}}$ ensures a large number of intervals over the full observation time, while $d$ itself cancels in the final sensitivity (Sec. \ref{sec: marginalizacion}).
Furthermore, we assume that each observation has constant variance $\epsilon^2$ and zero covariance.
In each interval of length $d$, the $\chi^2$ is minimized for the $\dot{n} d$ observations performed in that interval, obtaining an equation of the form
\begin{equation}
\frac{1}{\epsilon^2} \sum_{i=1}^{\dot{n} d} \bm{\mathsf{M}}^{i} \, \mathbf{\delta S} \;=\; \mathbf{D},
\end{equation}
where the sum runs over the observations in that interval, and $\mathbf{D}$ is the vector containing the data.

Approximating the summation by its mean value multiplied by the number of observations\footnote{This is justified by the condition $P_b \ll d$, which ensures that the orbital phase sweeps uniformly over many cycles within each interval.}, we obtain
\begin{equation}
\sum_{i=1}^{\dot{n} d} \bm{\mathsf{M}}^{i} \;\simeq\; \frac{\dot{n} d}{2\pi} \int_{0}^{2\pi} \bm{\mathsf{M}}^{i} \, d\Theta' 
\;=\; \dot{n} d \, \overline{\bm{\mathsf{M}}},
\end{equation}
where in the ELL1 model, $d\Theta'$ must be replaced by $d\Psi'$.
Finally, we obtain the covariance matrix:
\begin{equation}
    \bm{\mathsf{V}}= \frac{\epsilon^2}{\dot{n}  d} \overline{\bm{\mathsf{M}}}^{-1}, 
\end{equation}
whose diagonal elements are the variances of each parameter.

These variances were already obtained, for both the Blandford-Teukolsky and ELL1 models, in \cite{Kus2024}. Although that work studied a scalar dark matter field with linear coupling, because the residual function is identical, the parameter variances are also equal. This is because we are estimating the precision with which orbital parameters can be measured without introducing information about our specific dark matter model. The aforementioned results can be found in Appendix \ref{ap: varianzas}.

\subsection{Bayesian inference and marginalisation}
\label{sec: marginalizacion}

Assuming that the number of measurements $\mathcal{N} = T_{\rm obs}/d$ of the parameters is large and that they are taken regularly during the observation time $T_{\rm obs}$, we can use their variance to estimate the precision with which the model parameter perturbations can be measured. We are particularly interested in the perturbations induced by the presence of dark matter, as these measurements allow us to constrain the value of $\beta$ and $g$ over a range of masses.

To determine the sensitivity to $\beta$, we marginalise over the nuisance parameters considering the calculated variances. The variation of the parameters $\mathbf{S}$ over time, with their corresponding variances $\sigma_S^2$ as determined in Sec. \ref{sec: varianzas}, is given by:
\begin{subequations}
\begin{flalign}
    & \mathbf{\delta S}(t) = \mathbf{m}(\boldsymbol{\Xi},t) + \mathbf{h}(t) + \mathbf{n}, && \label{eq: senal completa} \\
    & \mathbf{h}(t) = \frac{\beta \, \Phi_{\mathrm{DM}}^2}{2} \, \left(\mathbf{A_X} \, X^2 + \mathbf{A_Y} \, Y^2 + \mathbf{A_{XY}} \, X \, Y \right) \label{eq: senal y descomposicion} && \\
    & \mathbf{m}(\boldsymbol{\Xi},t) = \boldsymbol{\Xi}^T \cdot N, &&
\end{flalign}
\end{subequations}
where $\Phi_{\text{DM}}=\frac{\sqrt{2 \rho_{\text{DM}}}}{m}$, $\boldsymbol{\Xi} = (\Xi_0, \Xi_1, \Xi_2)^T$ and $N = (1, t, t^2)^T$. The stochastic nature of the field is captured by the variables
\begin{equation}
    X = \sqrt{2} \,\varrho \, \cos \Upsilon, \quad Y = \sqrt{2} \, \varrho \, \sin \Upsilon.
\end{equation}

For instance, for the variable $x$, we have:
\begin{subequations}
\begin{align}
    A_X^{(x)}&= - x \left[\cos^2(mt) - \cos^2(m \, T_{\text{asc}})\right], \\
    A_Y^{(x)}&= - x \left[\sin^2(mt) - \sin^2(m \, T_{\text{asc}})\right], \\
    A_{XY}^{(x)}&= x \left[\sin(2mt) - \sin(2m \, T_{\text{asc}})\right],
\end{align}
\label{eq: descomposicion cuadratico x}
\end{subequations}
and for $\Psi'$ we find:
\begin{subequations}
\label{eq: descomposicion cuadratico psi}
\begin{align}
    A_X^{(\Psi')}(t) &= \frac{11}{16} \, \frac{\omega_b}{m} \left[ \sin(2m t) - \sin(2m T_{\mathrm{asc}}) \right], \label{eq:ax_psi_prima} \\
    A_Y^{(\Psi')}(t) &= -\frac{11}{16} \, \frac{\omega_b}{m} \left[ \sin(2m t) - \sin(2m T_{\mathrm{asc}}) \right], \label{eq:ay_psi_prima} \\
    A_{XY}^{(\Psi')}(t) &= \frac{11}{8} \, \frac{\omega_b}{m} \left[ \cos(2 m t) - \cos(2 m T_{\mathrm{asc}}) \right]. \label{eq:axy_psi_prima}
\end{align}
\end{subequations}

The component $\mathbf{m}$ in Eq. \eqref{eq: senal completa} includes constant, linear and quadratic in time corrections, where $\Xi_i$ are the constants that define the quadratic fit of \cite{BlandfordTeukolsky1976}. These terms capture corrections arising from physical effects other than ULDM, such as those included in Eq. \eqref{eq: cuadratico variacion psi} that have already been discussed. 

We define the Bayes factor as:
\begin{equation}
    \mathcal{B} = \frac{\mathcal{O}_h}{\mathcal{O}_n},
\end{equation}
where $\mathcal{O}_h$ is the evidence for the signal hypothesis and $\mathcal{O}_n$ is the evidence for the noise hypothesis. These are calculated according to:
\begin{equation}
    \mathcal{O}_i = \int d \vec{\lambda} \, \mathcal{L}_i (s, \vec{\lambda}_i) \, \Pi(\vec{\lambda}_i), 
\end{equation}
where $\vec{\lambda}$ is the vector of free parameters, $\mathcal{L}$ is the likelihood function, and $\Pi$ is the prior for the corresponding hypothesis. We will adhere to a Bayes factor of $\mathcal{B}=1000$ as a detection threshold, as done in \cite{Moore2015}. 

We assume that the noise $\mathbf{n}$ for each parameter is white and independent for each pulsar. Its probability distribution is given by:
\begin{equation}
    P(\mathbf{n})\,d\mathbf{n} = \frac{\exp\!\left(-\tfrac{1}{2}\,\mathbf{n}^T \mathbf{\Sigma_{n}^{-1}} \mathbf{n}\right)\,d\mathbf{n}}{\sqrt{(2\pi)^{\mathcal{N}} \det(\Sigma_{n})}},
    \label{eq: distrubucion ruido}
\end{equation}
where $\mathbf{\Sigma_{n}}$ is a diagonal matrix whose elements are the parameter variances $\sigma_S^2$. 

Substituting into the Rayleigh distribution of Eq. \eqref{eq: rayleigh} with the definitions of the new variables $X$ and $Y$, we obtain:
\begin{equation}
    \Pi(X,Y)= \frac{1}{2 \pi} e^{- \frac{X^2 + Y^2}{2}}.
    \label{eq: prior gaussiano}
\end{equation}
Thus, we have a Gaussian prior for the variables $X$ and $Y$.

Following \cite{Moore2015}, we further define the matrices
\begin{equation}
\mathbf{m} = \mathbf{M}\,\boldsymbol{\Xi}, 
\qquad 
\mathbf{M} \equiv \mathbf{U}\,\mathbf{T}\,\mathbf{V}^\dagger,
\qquad 
\mathbf{U} \equiv (\mathbf{F},\,\mathbf{G}),
\end{equation}
where $\mathbf{M}$ is the so-called design matrix, which admits a singular value decomposition into the matrices $\mathbf{U}$ (of dimensions $\mathcal{N} \times \mathcal{N}$), $\mathbf{T}$ (of dimensions $\mathcal{N} \times 3$), and $\mathbf{V}$ (of dimensions $3 \times 3$). 
Conveniently, $\mathbf{U}$ is written in terms of $\mathbf{F}$ and $\mathbf{G}$, with $\mathbf{G}$ being an $\mathcal{N} \times (\mathcal{N}-3)$ matrix. This matrix $\mathbf{G}$ is such that $\mathbf{G \, m^T}= \mathbf{0}$ and $\mathbf{G^T \, G}= \mathcal{I}_{\mathcal{N}-3}$.

Since we will adhere to a Bayes factor of 1000, we can assume that the posterior distributions of $m$ and $\beta$ (or $g$) will be strongly centered around their fiducial values. In consequence, for any reasonable prior, we can approximate it by
\begin{equation}
    \Pi(\beta,m) = \delta(\beta-\beta_f)\,\delta(m-m_f)
\end{equation}
or $\delta (g - g_f)$ in the case of vector ULDM. 

Regarding the treatment of the stochastic variables $X$ and $Y$, two distinct approaches may be followed to evaluate the Bayes factor. If we assume we know the specific realization of the ULDM field by approximating the priors as delta distributions,
\begin{equation}
    \Pi (X,Y) = \delta (X-X_f) \, \delta(Y-Y_f),
\end{equation}
the marginalisation becomes analytically trivial. This yields a closed-form expression for the noise-averaged Bayes factor, denoted by $\overline{\mathcal{B}}$:
\begin{equation}
    \ln (\overline{\mathcal{B}}) = \beta^2 \int_0^{T_{\text{obs}}} \frac{(G \, \tilde{h}_f^S)^2(t')}{2 \, \sigma_S^2\, d} dt' \equiv \beta^2 \, u_f^S,
    \label{eq: definicion integral u}
\end{equation}
where $h_f^S = \beta \, \tilde{h}_f^S$ represents the signal evaluated at the fiducial values, and $u_f^S$ corresponds to the marginalisation integral. The operator $G$, along with the relevant marginalisation integrals of this section, is described in Appendix \ref{ap:marg}.

Lastly, we can combine the results of different pulsars to obtain a combined sensitivity curve: 
\begin{equation}
    \overline{\mathcal{B}}_C= \prod_{p=1}^{N_p} \overline{\mathcal{B}}_p,
    \label{eq: bayes combinado}
\end{equation}
where $\overline{\mathcal{B}}_C$ is the combined Bayes factor, $ \overline{\mathcal{B}}_p$ is the Bayes factor for an individual pulsar and $N_p$ is the number of systems. 

Since $\overline{\mathcal{B}}_p= \exp(\beta^2 \, u_{fp}^S)$,
\begin{equation}
    \beta = \sqrt{\frac{\ln (\overline{\mathcal{B}}_C)}{\sum_p  u_{fp}^S}}
\end{equation}

As for the case of spin-1 ULDM, the results are identical, making the substitution $\beta \rightarrow g$. Note that this two-step method allows us to compute the sensitivity to one orbital parameter $S$ at a time. 

If we use the Gaussian prior given by Eq. \eqref{eq: prior gaussiano} instead, the expression for the Bayes factor for quadratically coupled scalar dark matter is
\begin{equation}
\begin{aligned}
\overline{\mathcal{B}} = \Biggl[ &(1 - \beta^2 \Phi_{\text{DM}}^2 u_x^S)(1 - \beta^2 \Phi_{\text{DM}}^2 \, u_y^S) \\
&- \left(\frac{\beta^2 \Phi_{\text{DM}}^2}{2}\, u_{xy}^S\right)^2 \Biggr]^{-1/2},
\end{aligned}
\label{eq: resultado complicado factor de bayes}
\end{equation}
where $u_x^S$, $u_y^S$ and $u_{xy}^S$ are defined in Eq. \eqref{eq: marginalisation integrals} in Appendix \ref{ap:marg}.

While it is possible to set $\overline{\mathcal{B}}=1000$ and solve Eq. \eqref{eq: resultado complicado factor de bayes} for an individual pulsar, in order to combine different systems the polynomial equation of degree $2 \, N_p$ that arises from Eq. \eqref{eq: bayes combinado} was solved numerically. 

\section{Sensitivity estimates}
\label{sec:results}

\subsection{Spin-0 ultralight dark matter}
\label{sec: results spin0}

In this section, we present the sensitivity curves obtained by combining binary systems in circular orbits (modeled via the ELL1 framework) and eccentric systems exhibiting resonances (through the BT model). We utilize the current observational precision from NANOGrav \cite{NANOGrav2023}, summarized in Appendix \ref{ap: tablas}.

It is important to emphasize that the following results constitute a sensitivity estimation rather than a direct data analysis of timing residuals. Since the local field parameters $\varrho$ and $\Upsilon$ are inherently stochastic and unknown for each pulsar system, we sample these variables from their underlying probability distributions. Each combined curve thus represents a single realization of the global dark matter field for each binary system; consequently, an ensemble of such realizations is plotted to characterize the statistical variation.

For circular orbits we consider the systems presented in Table \ref{tab_pulsar_data}. We focus on the parameters $x$ and $\Psi'$, which were identified as the most sensitive observables in the study of universal linear coupling presented in \cite{Kus2024}. Deriving the sensitivity curves reduces to evaluating the integrals $u_x^S$, $u_y^S$, and $u_{xy}^S$ over the observation time for the Gaussian prior case, and the integral $u_f^S$ when assuming delta priors for the variables $X$ and $Y$. The sensitivity curve $\beta(m)$ is defined by requiring a Bayes factor threshold of 1000 as the criterion for detection.

\begin{figure}[t!]
    \centering
    \includegraphics[width=0.99\linewidth]{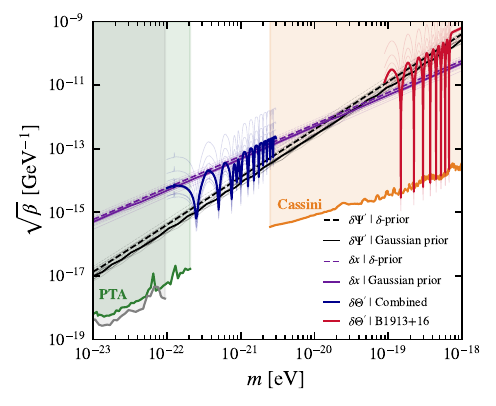}
    \caption{Sensitivity for quadratically coupled scalar ULDM derived from binary pulsar timing. In all cases, solid lines represent the ensemble medians, while the lighter-colored curves depict the individual stochastic field realizations. The blue curve corresponds to the limits for three eccentric systems (PSR J1903+0327, PSR J1946+3417 and PSR J2234+0611) derived via the parameter $\Theta'$, while the red curve shows the result for the eccentric PSR B1913+16. The black and purple curves exhibit the results for systems in circular orbits, obtained with $\Psi'$ and $x$ respectively. In the case of resonant systems, marginalisation was only performed over $\Xi_0$ and $\Xi_1$. Shaded regions indicate the parameter space excluded by Cassini \cite{Armstrong:2003hb} (orange) and Pulsar Timing Array \cite{Porayko:2018sfa, Smarra:2023rjx} (green and grey, respectively) observations.}
    \label{fig: resultados cuadratico}
\end{figure}

Fig. \ref{fig: resultados cuadratico} illustrates the results for both parameters, considering Gaussian and delta priors across ten realizations of the stochastic variables $\Upsilon$ and $\varrho$ for each case. Notably, the choice of prior does not have a significant impact on the sensitivity curves, resulting only in order-of-unity differences. In fact, the curves corresponding to both prior types largely overlap, occupying approximately the same region of the parameter space. Furthermore, it is observed that for $m \lesssim 2 \times 10^{-20}\,\mathrm{eV}$, the parameter $\Psi'$ is more sensitive to the presence of dark matter, whereas $x$ imposes more restrictive constraints at higher masses.

In the case of eccentric orbits, we adopted delta priors for both theoretical and computational simplicity; given the results for circular orbits, no appreciable differences are expected compared to the use of a Gaussian prior. We employed the subset of NANOGrav-measured systems from Table \ref{tab:pulsars_high_eccentricity} to construct a combined sensitivity curve based on the parameter $\Theta'$. The Hulse–Taylor binary PSR B1913+16 is treated separately and shown as its own curve, since its substantially shorter orbital period places its resonant masses ($m= \frac{n \, \omega_b}{2}$) in a higher-mass window than that probed by the NANOGrav eccentric millisecond pulsars. The substantially longer observation span of PSR B1913+16 (over three decades, compared to $\sim 10$ years for the NANOGrav systems) is also reflected in the morphology of its resonant peaks. As shown in \cite{blas2019}, the on-resonance sensitivity scales as $T_{\rm obs}^{-5/2}$, while the off-resonance sensitivity scales only as $T_{\rm obs}^{-1/2}$. Consequently, the resonant peaks become sharper and deeper as $T_{\rm obs}$ increases. This explains the markedly more pronounced resonant features visible in the PSR B1913+16 curve relative to those of the eccentric millisecond pulsars.

Following \cite{Kus2024}, we have marginalised over $\Xi_0$ and $\Xi_1$ but not the quadratic coefficient. The difference between including and neglecting this contribution is only noticeable in a narrow mass range near resonance, since in this limit the resonant secular drift becomes degenerate with the $\Xi_2$ term. We therefore assume that $\Xi_2$ is either negligible with respect to the ULDM resonant effect or that it is known and subtracted so that it can be set to zero. 

It is worth highlighting that the combined curve obtained from $\delta\Psi'$ generally imposes stricter constraints on the coupling constant than the combination of the three eccentric systems, except in the vicinity of resonant masses. Near these resonances, the curves reproduce the limits obtained in \cite{Blas2017Resonates} in order of magnitude, where the system response is strongly amplified as the dark matter frequency coincides with a harmonic of the orbital frequency. Our method therefore yields a continuous coverage that bridges the gaps inaccessible to resonant-regime analyses.

Outside these resonant regions, the combination of circular-orbit systems proves more efficient, yielding more restrictive bounds on the coupling constant. In this sense, the main advantage of the method lies in the complementarity between the two approaches: eccentric systems dominate sensitivity near resonances, while the combination of circular systems allows for stricter limits throughout the remainder of the mass range.

We observe that both the Cassini and Pulsar Timing Array\footnote{The Cassini \cite{Armstrong:2003hb} and PTA \cite{Porayko:2018sfa, Smarra:2023rjx} bounds are obtained by recasting their reported amplitude limits onto our coupling as explained in \cite{Blas2017Resonates}.} constraints are several orders of magnitude more restrictive than those obtained in this work within their respective mass ranges. A notable exception is found at the resonant peaks of PSR B1913+16, where our bound approaches the Cassini limit. However, our method allows us to constrain the coupling constant in the mass interval between $2 \times 10^{-22}\,\mathrm{eV}$ and $2 \times 10^{-21}\,\mathrm{eV}$, which is inaccessible to those experiments. In this regard, our results provide an independent and complementary bound for quadratically coupled ULDM. 

A final caveat concerns the interpretation of our bounds. A quadratically coupled scalar acquires an effective mass $m_{\rm eff}^2 \simeq  m^2_{\Phi}+\beta (\rho-3 p)$ inside a body of density $\rho$ and pressure $p$, so that  depending on the coupling and the  equation of state of the neutron star there might be   significant screening  of the local  field leading to a lower effective coupling. As studied  in Refs. \cite{T_Damour_1992, KuntzBarausse} and mentioned in  \cite{blas2019, Blas2017Resonates}  this screening is expected to be operative for neutron stars throughout the parameter space probed here, and may degrade the bounds on ${\beta}$ by a few orders of magnitude with respect to those displayed.


Non-perturbative phenomena can alter the response of a neutron star more dramatically; these typically require large negative $\beta$ \cite{blas2019}, and we take $\beta > 0$ throughout. Quantifying either effect would require a model of the neutron star interior for a given equation of state, along the lines of the model used to  compute the  sensitivities in \cite{KuntzBarausse} and applied to pulsar timing data as in \cite{smarra2024constraints}, which lies beyond our scope. Our bounds are therefore expected to be representative at the order-of-magnitude level.

\subsection{Spin-1 ultralight dark matter}
\label{sec: results spin1}

As in the scalar case with quadratic coupling, the sensitivity curves for vector ULDM are obtained by applying the same statistical framework and multi-system combination process. In this instance, however, the time dependence of $\delta\Psi'(t)$ and $\delta\Theta'(t)$ is considerably more complex, making a closed-form analytical expression for the full signal impractical. Consequently, the signal associated with the variation of these variables is constructed through direct numerical integration of the orbital parameter time derivatives, according to Eqs. \eqref{eq: delta theta' spin 1} and \eqref{eq: delta_Psi_general_spin1}. In addition to the stochastic field amplitude $\varrho$ and phase $\Upsilon$, the vector case requires sampling the two angles $\vartheta$ and $\varphi$ that specify the field direction in Eq. \eqref{eq: vector field}. While $\varphi$ is drawn uniformly in $[0, 2\pi)$, a naive uniform sampling of $\vartheta$ in $[0, \pi]$ would oversample field orientations near the poles relative to the equator and therefore bias the sensitivity estimates. Following \cite{DeepNeuralULDM}, we instead sample $\cos\vartheta$ uniformly in $[-1,1]$, ensuring an isotropic distribution of field directions over the sphere.

\begin{figure}[t!]
    \centering
    \includegraphics[width=0.99\linewidth]{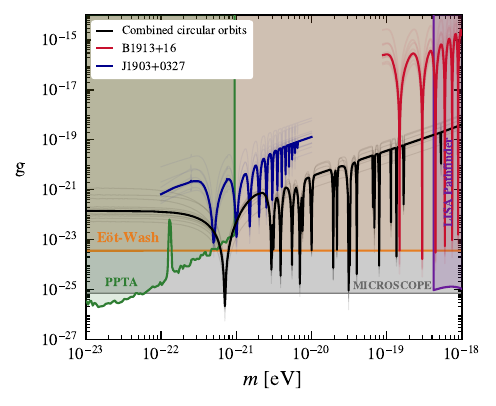}
    \caption{Constraints on the vector ULDM coupling strength $g$ as a function of mass. In all cases, solid lines represent the median sensitivity, while the lighter individual field realizations are computed assuming delta priors for the stochastic parameters. The black curves correspond to the combined analysis of 23 circular binary systems, utilizing the two-step method applied to the $\Psi'$ variable. For the eccentric systems, the blue and red curves represent the bounds derived from PSR J1903+0327 and PSR B1913+16, respectively, applying our method to the $\Theta'$ variable. For both circular and eccentric orbits we have marginalised over $\Xi_0$ and $\Xi_1$, as done in the scalar case. Shaded regions indicate the parameter space excluded by MICROSCOPE \cite{Amaral:2024vwd} (grey), Eöt-Wash \cite{Shaw:2021tb} (orange), LISA Pathfinder \cite{Frerick:2023dm} (purple), and the Parkes Pulsar Timing Array \cite{Xue:2021dp} (green).}
    \label{fig: resultados spin 1}
\end{figure}

In Fig. \ref{fig: resultados spin 1}, we present the upper bounds for the coupling constant $g$ obtained by combining twenty-three binary pulsars in near-circular orbits, assuming $\Delta c=0.1$. Each peak in the curve corresponds to a resonance between the dark matter mass and a system's orbital frequency. In addition to the circular orbit results, we derive sensitivity curves for PSR J1903+0327 and PSR B1913+16, identified in \cite{spin1} as the most sensitive eccentric systems. As in the scalar case, these two pulsars are studied separately due to their orbital periods placing their resonances in disjoint mass windows. In all cases, the on-resonance sensitivity for these systems is consistent with the levels reported in the aforementioned work. We find that the combined curve for circular systems generally provides more stringent constraints on $g$ across most of the mass range, except in the immediate vicinity of PSR B1913+16’s orbital harmonics. This result is expected; since circular orbits also exhibit resonant behavior, the statistical combination of a larger number of systems leads to a significant improvement in the overall bounds.

Our results are presented alongside constraints from various precision tests of the weak equivalence principle, specifically targeting B-L spin-1 couplings. These include MICROSCOPE limits based on differential acceleration measurements aboard a drag-free satellite \cite{Amaral:2024vwd} and Eöt-Wash torsion balance experiments designed to detect periodic torques \cite{Shaw:2021tb}. Furthermore, we include reinterpretations of LISA Pathfinder data \cite{Frerick:2023dm} and Parkes Pulsar Timing Array searches for stochastic gravitational waves \cite{Xue:2021dp}. 

Comparison with these precision experiments shows that the sensitivity achieved through our proposed method is within orders of magnitude of current state-of-the-art bounds. Although MICROSCOPE and Eöt-Wash provide stricter limits over most of the analyzed mass range, our sensitivity exceeds the Eöt-Wash bounds near the orbital resonances of the circular systems as well as near the first three harmonics of PSR B1913+16. In these resonant regions, our constraints are outperformed only by MICROSCOPE ---except in a narrow mass window near the orbital resonance of PSR J1713+0747 ($P_b=67.83$ d, $m\sim 7 \times 10 ^{-22}$ eV), where the combined circular-orbit bound surpasses even that limit. 

Beyond the comparison with current bounds, a fundamental advantage of our approach lies in its scalability. The sensitivity can be significantly enhanced by increasing the number of combined pulsars and reducing the white noise associated with TOA measurements, as expected with the advent of next-generation radio observatories. Since the combined sensitivity scales as $N_p^{-1/2}$ for systems of comparable precision, monitoring of order $ \sim 10^3$ pulsars with the Square Kilometre Array would translate into a more than one-order-of-magnitude improvement on $g$ relative to the bounds presented here.

\section{Conclusions}
\label{sec:conclusions}

In this work, we implemented a theoretical and statistical framework to quantify the sensitivity of binary pulsars as detectors of ultralight dark matter. By applying pulsar timing formalisms, we modeled the perturbations induced by ULDM fields on pulse times of arrival. This paper focused on both quadratically coupled spin-0 scalar fields and spin-1 vector fields, allowing us to translate dynamical orbital perturbations into observational constraints on their respective coupling constants using data on the observational precision for real binary systems.

The employed two-step Bayesian approach, centered on the explicit marginalisation over nuisance orbital parameters, provides a statistically consistent description that accounts for the inherent uncertainties in the binary model. This ensures that the resulting sensitivity limits are realistic, avoiding the artificial overestimation that can arise from simpler fitting procedures as seen in Appendix \ref{ap:one-step}. Furthermore, the framework is robust against the choice of dynamical variables ---whether the angle $\Theta'$ or the pulsar's true anomaly is chosen, as discussed in Appendix \ref{ap: anomalia}.

For the case of quadratically coupled scalar ULDM, we obtained continuous sensitivity curves across a wide range of masses, covering regions of the parameter space that are currently challenging for other astrophysical probes. We identified regions where different orbital parameters dominate the sensitivity: while the orbital phase $\Psi'$ is more sensitive at lower masses, the projected semi-major axis $x$ becomes more restrictive at higher masses. Our results are particularly competitive in the mass range between $2\times 10^{-22}\,\mathrm{eV}$ and $2\times 10^{-21}\,\mathrm{eV}$, naturally complementing the limits imposed by the Cassini mission and Pulsar Timing Array observations. The complementarity between circular and eccentric systems plays a central role. While combined circular orbits dominate the sensitivity throughout most of the mass range, the eccentric systems provide stronger constraints in the vicinity of resonant masses; in particular, PSR B1913+16, whose long observation span sharpens its resonant features, approaches the Cassini limit at its resonant peaks.

Regarding vector ULDM, we confirmed that direct interaction with binary systems induces resonant signatures that remain present even in nearly circular orbits. By combining the timing data of twenty-three binary systems, we derived constraints on the vector coupling constant $g$ that surpass those of individual eccentric systems across most of the mass range and reach orders of magnitude comparable to the constraints derived from high-precision laboratory and space-based experiments, such as MICROSCOPE and Eöt-Wash. Notably, our combined circular-orbit bound surpasses MICROSCOPE in a narrow mass window near the orbital resonance of PSR J1713+0747, illustrating that pulsar-based searches can already be competitive with state-of-the-art equivalence-principle tests in specific regions of parameter space.

The current framework would benefit from a direct analysis of real time-of-arrival data, complemented by simulated datasets, to further validate the robustness of the statistical limits and the impact of realistic noise profiles. Looking forward, future observation campaigns, such as those planned with the Square Kilometre Array, will significantly increase the number of available binary pulsars. Considering that the method presented here allows for the seamless combination of multiple systems, we expect that the constraints on $\beta$ and $g$ can be improved by at least one order of magnitude. This advancement will position pulsar astronomy as one of the most powerful tools for exploring dark matter physics in the ultralight regime, both by tightening the bounds on these couplings with larger datasets and by extending the formalism to more complex scenarios, such as tensor ULDM fields
or non-universal couplings.

In the quadratic scalar case, non-universality is induced by screening. A future analysis could integrate both this effect and the modulation of the pulsar spin frequency arising from the same variation of the constituent masses, which would require a model of the stellar interior for a given equation of state.

\section*{Data availability statement}
The data and numerical implementation that support the findings of this article are openly available in the GitHub repository \cite{codigo_usado_2026}, which includes the Python scripts and processed results necessary to reproduce the sensitivity curves presented. 

\section*{Acknowledgments}
This work has been supported by CONICET and UBA. We thank Aurélien Hees for inputs on constraints from alternative  experiments.  We acknowledge Daphne Estrada for analytic calculations in the quadratic case, carried out during her MSc thesis. 

\appendix
\section{Osculating equations}
\label{ap: perturbed orbital parameters}

We employ six orbital parameters to describe a Newtonian orbit \cite{turner1979}: 
\begin{subequations}
\label{eq:gauss_equations_left}
\begin{flalign}
    & \dot{a} = \frac{2}{\omega_b} \left\{ \frac{F_r \, e}{\sqrt{1 - e^2}} \sin\theta + \frac{F_\theta}{r} \, a \, \sqrt{1 - e^2} \right\} && \label{eq:gauss_a} \\
    & \dot{e} = \frac{\sqrt{1 - e^2}}{a \, \omega_b} \left\{ F_r \sin\theta + F_\theta \left( \cos\theta + \frac{1}{e} - \frac{r}{ae}\right) \right\} && \label{eq:gauss_e} \\
    & \dot{\Omega} = \frac{(1 - e^2)^{-1/2} \, r}{a^2 \omega_b} \, \frac{F_z \sin(\theta + \omega)}{\sin \iota} && \label{eq:gauss_Omega} \\
    & \dot{\iota} = \frac{F_z \, r \, \cos(\theta + \omega)}{a^2 \omega_b \sqrt{1 - e^2}} && \label{eq:gauss_iota} \\
    & \dot{\varpi} = \frac{\sqrt{1 - e^2}}{a \,e \, \omega_b} \left\{ - F_r \cos\theta + F_\theta \sin\theta \left[1 + \frac{r}{a(1 - e^2)} \right] \right\} \nonumber \\
    & \hspace{25pt} + 2 \sin^2 \left( \frac{\iota}{2} \right) \dot{\Omega} && \label{eq:gauss_varpi} \\
    & \dot{\epsilon}_1 = -\frac{2 r}{a^2 \omega_b} F_r + \left( 1 - \sqrt{1 - e^2}\right) \, \dot{\varpi} \nonumber \\
    & \hspace{25pt} + 2 \sqrt{1 - e^2} \, \sin^2\left(\frac{\iota}{2}\right) \dot{\Omega} && \label{eq:gauss_epsilon}
\end{flalign}
\end{subequations}
\noindent with
$\varpi = \omega + \Omega$, $\epsilon_1 = \omega_b (t - T) + \varpi - \int \omega_b \, dt$, and $\cos E = \frac{1}{e} \left(1 - \frac{r}{a} \right)$.

The orbital motion in eccentric systems lacks a closed-form analytical expression. Consequently, the dynamics are typically represented through a Fourier series decomposition of unperturbed Keplerian orbits, which is equivalent to an expansion of the form \cite{bessel}:
\begin{subequations}
\begin{align}
\frac{r}{a} &= 1 + \frac{e^2}{2} - 2e \sum_{n=1}^{\infty} \frac{J_n'(ne)}{n} \cos(n \omega_b \tilde{t}) \,,  \\[1ex]
\frac{r^2}{a^2} &= 1 + \frac{3}{2} e^2 - 4 \sum_{n=1}^{\infty} \frac{J_n(ne)}{n^2} \cos(n \omega_b \tilde{t}) \,,  \\[1ex]
\cos\theta &= -e + \frac{2(1 - e^2)}{e} \sum_{n=1}^{\infty} J_n(ne) \cos(n \omega_b \tilde{t}) \,,  \\[1ex]
\sin\theta &= 2 \sqrt{1 - e^2} \sum_{n=1}^{\infty} J_n'(ne) \sin(n \omega_b \tilde{t}) \,,  \\[1ex]
\frac{r}{a} \cos\theta &= -\frac{3e}{2} + 2 \sum_{n=1}^{\infty} \frac{J_n'(ne)}{n} \cos(n \omega_b \tilde{t}) \,,  \\[1ex]
\frac{r}{a} \sin\theta &= \frac{2 \sqrt{1 - e^2}}{e} \sum_{n=1}^{\infty} \frac{J_n(ne)}{n} \sin(n \omega_b \tilde{t}) \,, \\[1ex]
\frac{a^2}{r^2} \cos\theta &= \sum_{n=1}^{\infty} 2n \,  J_n'(ne) \cos(n \omega_b \tilde{t}) \,,  \\[1ex]
\frac{a^2}{r^2} \sin\theta &= \frac{\sqrt{1 - e^2}}{e} \sum_{n=1}^{\infty} 2n \, J_n(ne) \sin(n \omega_b \tilde{t}) \,, 
\end{align}
\end{subequations}


\noindent where $\tilde{t} = t - T_0$, $J_n(x)$ is the Bessel function of order $n$, and $J_n'(x)$ its derivative. An alternative treatment of these dynamics, based on the angle-action formalism, was developed in \cite{Desjacques_2020} for the case of an oscillating axion background.
\onecolumngrid
\section{Perturbed orbital parameters}
\label{ap: orbital perturb}
In this Appendix, we collect the expressions obtained from integrating the osculating equations for the different Keplerian parameters in the case of scalar ULDM. All the resonant contributions can be written in terms of the single function
\begin{equation}
\mathcal{C}^{\pm}_n(t) \equiv
\frac{\cos\Upsilon_0-\cos\!\big[(n\omega_b\pm 2m)\,\tilde t\pm\Upsilon_0\big]}
     {n\omega_b\pm 2m},
\end{equation}
where $\tilde t = t-T_0$ and $\Upsilon_0 \equiv 2mT_0+2\Upsilon$. With this notation,
\begin{align}
\frac{\delta a}{a} &= -2\beta\Phi_0^2\sum_n J_n(ne)
\Bigg[\frac{1-\cos(n\omega_b\tilde t)}{2}
+\Big(\frac{n\omega_b}{4}-m\Big)\mathcal{C}^{+}_n
+\Big(\frac{n\omega_b}{4}+m\Big)\mathcal{C}^{-}_n\Bigg]
\\
&\quad-\beta\Phi_0^2\Big[\cos^2(mt+\Upsilon)-\cos^2(mT_0+\Upsilon)\Big]
+\frac{\delta a_0}{a}+\frac{\dot a_0\,\tilde t}{a},
\nonumber \\[6pt]
\delta e(t) &= \beta\Phi_0^2\,\frac{1-e^2}{e}\sum_n J_n(ne)
\Bigg[-\frac{1-\cos(n\omega_b\tilde t)}{2}
-\Big(\frac{n\omega_b}{4}-m\Big)\mathcal{C}^{+}_n
-\Big(\frac{n\omega_b}{4}+m\Big)\mathcal{C}^{-}_n\Bigg]
+\delta e_0+\dot e_0\,\tilde t,
\\[6pt]
\delta\omega(t) &= \beta\Phi_0^2\,\frac{\sqrt{1-e^2}}{e}\sum_n J_n'(ne)
\Bigg[\frac{\sin(n\omega_b t)-\sin(n\omega_b T_0)}{2}
+\Big(\frac{n\omega_b}{4}-m\Big)\big(\mathcal{C}^{+}_n+\mathcal{C}^{-}_n\big)\Bigg]
+\delta\omega_0+\dot\omega_0\,\tilde t .
\end{align}
\twocolumngrid
\section{Variances}
\label{ap: varianzas}
The time residual corresponding to the Blandford--Teukolsky model can be expressed in terms of six independent parameters:
\(\delta K\), \(\delta a\), \(\delta e\), \(\delta \alpha_b\), \(\delta \eta_b\) and \(\delta \Theta'\).
The explicit dependence on \(\delta a\) arises solely from the appearance of the semi-major axis in the last term of the BT timing model, which constitutes a subdominant correction. In what follows, and following the original treatment of \cite{BlandfordTeukolsky1976}, this contribution is neglected when estimating the variances of the remaining parameters.

Under this approximation, the analysis is restricted to the set
\(\{\delta K,\, \delta e,\, \delta \alpha_b,\, \delta \eta_b,\, \delta \Theta'\}\),
and the dominant contribution to the time residual can be written as
\begin{equation}
\begin{aligned}
R^{\mathrm{BT}} =&
\, \, \delta \eta_b \, \sin E'
 +\delta \alpha_b (\cos E' - e)
\\&-\frac{\delta \Theta'\,(\alpha_b \sin E' - \eta_b \cos E')}{1 - e \cos E'} \, 
\\&-\delta e \left[
\alpha_b
+\frac{\sin E'\,(\alpha_b \sin E' - \eta_b \cos E')}{1 - e \cos E'}
\right].
\end{aligned}
\label{eq:BT_residual_dominant}
\end{equation}
From this expression, the variances of the parameters, for generic values of $e$, $\alpha_b$ and $\eta_b$, are given by
\par\medskip\begin{widetext}
\begin{subequations}
\label{eq:variances_subeqs}
\begin{align}
    \begin{split}
        \mathrm{var}(\delta K) ={} & - Q \Big[ \eta_b^4 \big( e^4 (26 - 9\schwa) + 4 e^2 (8\schwa - 7) + 8(\schwa - 1) \big) \\
        & + 2 \alpha_b^2 \eta_b^2 \big( e^6 + e^4 (18 - 5\schwa) + 4 e^2 (6\schwa - 5) + 8(\schwa - 1) \big) \\
        & + \alpha_b^4 \big( 8 e^8 + e^6 (32\schwa - 62) + e^4 (74 - 65\schwa) + 4 e^2 (4\schwa - 3) + 8(\schwa - 1) \big) \Big], 
    \end{split} \label{eq:var_deltaK} \\[1ex]
    \begin{split}
        \mathrm{var}(\delta \alpha_b) ={} & Q \Big[ 4 \eta_b^4 \big( e^2 (\schwa - 3) - 4\schwa + 4 \big) \\
        & + 2 \alpha_b^2 \eta_b^2 \big( 7 e^4 - 22 e^2 - 2 (e^4 - 7 e^2 + 8)\schwa + 16 \big) \\
        & + \alpha_b^4 \big( e^6 (\schwa - 6) + e^4 (30 - 16\schwa) + 8 e^2 (4\schwa - 5) - 16(\schwa - 1) \big) \Big], 
    \end{split} \label{eq:var_deltaalphab} \\[1ex]
    \begin{split}
        \mathrm{var}(\delta \eta_b) ={} & Q \Big[ 4 \alpha_b^4 (e^2 - 1)^2 \big( e^2 (\schwa - 3) - 4\schwa + 4 \big) \\
        & + 2 \eta_b^4 \big( e^4 + 4 e^2 (\schwa - 2) - 8\schwa + 8 \big) \\
        & + \alpha_b^2 \eta_b^2 \big( e^6 (\schwa - 6) + e^4 (42 - 20\schwa) + e^2 (52\schwa - 68) - 32(\schwa - 1) \big) \Big], 
    \end{split} \label{eq:var_deltaetab} \\[1ex]
    \mathrm{var}(\delta e) ={} & Q e^4 \Big[ \alpha_b^2 (e^2 - 2)\big( e^2 (\schwa - 2) - 2\schwa + 2 \big) - 2 \eta_b^2 (e^2 - 2)\big( e^2 + 2\schwa - 2 \big) \Big], \label{eq:var_deltae} \\[1ex]
    \mathrm{var}(\delta \Theta') ={} & Q e^2 \schwa \Big[ 2 \alpha_b^2 (e^2 - 1)\big( e^2 (\schwa - 2) - 2\schwa + 2 \big) + \eta_b^2 (e^2 - 2)(e^2 + 2\schwa - 2) \Big]. \label{eq:var_deltaTheta}
\end{align}
\end{subequations}
where
\begin{equation*}
    Q = \frac{\epsilon^2}{ d\,\dot{n}\, \big[ e^4 + 4 e^2 (\schwa - 2) - 8\schwa + 8 \big]\, \big[ \eta_b^2 - \alpha_b^2 (e^2 - 1) \big]^2},
\end{equation*}
\end{widetext}
and we have defined $\schwa \equiv \sqrt{1 - e^2}$. 

Within the ELL1 formalism it is possible to obtain analytical expressions for the variances associated with the six independent orbital parameters. However, these general expressions are lengthy and not particularly transparent. For this reason, and following the standard treatment, we present here only the results corresponding to the case $\eta = \kappa = 0$, which captures the relevant behavior in the small-eccentricity regime.

In this limit, the variances are given by
\begin{subequations}
\label{eq:ELL1_variances}
\begin{align}
    \mathrm{var}(\delta K) &= \frac{\epsilon^2\!\left(34 x^4 \omega_b^4 + 61 x^2 \omega_b^2 + 76\right)} {d\,\dot{n}\!\left(34 x^4 \omega_b^4 - 11 x^2 \omega_b^2 + 4\right)}, \label{eq:ELL1_var_K} \\
    \mathrm{var}(\delta s) &= \frac{4 s^2 \epsilon^2\!\left(5 x^4 \omega_b^4 + 80 x^2 \omega_b^2 + 32\right)} {81\, d\,\dot{n}\, x^6 \omega_b^4}, \label{eq:ELL1_var_s} \\
    \mathrm{var}(\delta x) &= \frac{20\,\epsilon^2}{9\, d\,\dot{n}}, \label{eq:ELL1_var_x} \\
    \mathrm{var}(\delta \eta) &= \frac{32\,\epsilon^2\!\left(x^2 \omega_b^2 + 1\right)} {d\,\dot{n}\, x^2\!\left(34 x^4 \omega_b^4 - 11 x^2 \omega_b^2 + 4\right)}, \label{eq:ELL1_var_eta} \\
    \mathrm{var}(\delta \kappa) &= \frac{32\,\epsilon^2}{9\, d\,\dot{n}\, x^4 \omega_b^2}, \label{eq:ELL1_var_kappa} \\
    \mathrm{var}(\delta \Psi') &= \frac{4\,\epsilon^2\!\left(17 x^2 \omega_b^2 + 2\right)} {d\,\dot{n}\, x^2\!\left(34 x^4 \omega_b^4 - 11 x^2 \omega_b^2 + 4\right)}  \label{eq:ELL1_var_Psi}
\end{align}
\end{subequations}

These expressions can be further simplified by noting that the combination
\begin{equation}
\omega_b^2 x^2
=
\frac{G M_2}{a}\,
\frac{\sin^2\iota}{1 + M_1/M_2}
\label{eq:omega_x_small}
\end{equation}
is typically much smaller than unity for binary pulsar systems. Consequently, retaining only the leading terms in the limit $\omega_b^2 x^2 \ll 1$, one obtains the approximations
\begin{subequations}
\label{eq:ELL1_var_leading}
\begin{align}
    \mathrm{var}(\delta K) &\simeq \frac{19\,\epsilon^2}{d\,\dot{n}}, \label{eq:ELL1_var_leading_K} \\
    \mathrm{var}(\delta s) &\simeq \frac{128\, s^2 \epsilon^2}{81\, d\,\dot{n}\, x^6 \omega_b^4}, \label{eq:ELL1_var_leading_s} \\
    \mathrm{var}(\delta x) &\simeq \frac{20\,\epsilon^2}{9\, d\,\dot{n}}, \label{eq:ELL1_var_leading_x} \\
    \mathrm{var}(\delta \eta) &\simeq \frac{8\,\epsilon^2}{d\,\dot{n}\, x^2}, \label{eq:ELL1_var_leading_eta} \\
    \mathrm{var}(\delta \Psi') &\simeq \frac{2\,\epsilon^2}{d\,\dot{n}\, x^2}, \label{eq:ELL1_var_leading_Psi} \\
    \mathrm{var}(\delta \kappa) &\simeq \frac{32\,\epsilon^2}{9\, d\,\dot{n}\, x^4 \omega_b^2}. \label{eq:ELL1_var_leading_kappa}
\end{align}
\end{subequations}
\section{Marginalisation integrals}
\label{ap:marg}

Here, we collect the integrals required for marginalisation over the nuisance parameters. Firstly, we define the $G$ operator, which projects the signal on the subspace orthogonal to polynomials of up to degree 2 in time: 
\begin{equation}
    (Gf)(t)=f(t) - \sum_{i= m_s}^2 f_i(t) \, \frac{\int_0^{T_{obs}}f_i(\tau) \, f(\tau) \, d\tau}{\int_0^{T_{obs}}f_i^2(\tau) \, d \tau},   
\end{equation}
where $f_i(\tau)$ are
\begin{subequations}
\begin{align}
    f_0(\tau) &= \frac{\tau^2}{T_{obs}} - \tau + \frac{T_{obs}}{6}, \\
    f_1(\tau) &= \tau - \frac{T_{obs}}{2}, \\
    f_2(\tau) &= T_{obs};
\end{align}
\end{subequations}
also, $m_s=0$ if $\Xi_2 \neq 0$, $m_s=1$ if $\Xi_2 = 0$ but $\Xi_1 \neq 0$, and $m_s=2$ if $\Xi_2 = \Xi_1=0$ but $\Xi_0 \neq 0$.

As for the marginalisation integrals introduced in section \ref{sec: marginalizacion}, they are defined as
\begin{subequations}
\begin{align}
    u_x^S &= \int_0^{T_{obs}} \, \frac{(G \, A_X^S)(t') \, (G \, \tilde{h}_f^S)(t')}{2 \, \sigma_S^2 \, d} dt' \\
    u_y^S &= \int_0^{T_{obs}} \, \frac{(G \, A_Y^S)(t') \, (G \, \tilde{h}_f^S)(t')}{2 \, \sigma_S^2 \, d} dt' \\
    u_{xy}^S &= \int_0^{T_{obs}} \, \frac{(G \, A_{XY}^S)(t') \, (G \, \tilde{h}_f^S)(t')}{2 \, \sigma_S^2 \, d} dt'.
\end{align}
\label{eq: marginalisation integrals}
\end{subequations}
\section{Observational precision}
\label{ap: tablas}

In Table \ref{tab_pulsar_data}, we include the data reported by the NANOGrav 15-year release \cite{NANOGrav2023} for the different pulsars in near-circular orbits studied in this work, described by the ELL1 timing model. In cases where the pulsar and companion masses were not reported, we assumed an orbital inclination of $\iota = 60^{\circ}$ ---since it represents the median value for a random distribution of orbital orientations in 3D space---, a pulsar mass of $1.35 \, M_{\odot}$, and adopted the companion mass reported in \cite{ATNF2025}. The exceptionally high companion mass of $2.97 \, M_{\odot}$ reported for PSR J0613-0200 corresponds to a Shapiro delay measurement with large uncertainties, and should be interpreted with caution.

Table \ref{tab:pulsars_high_eccentricity} summarizes the orbital and observational parameters for the high-eccentricity binary pulsars analyzed in this study. In addition to the variables listed in Table \ref{tab_pulsar_data}, we include the argument of periastron ($\omega$), as it enters the osculating equations for vector ULDM in eccentric orbits. Unlike the other systems, PSR B1913+16 was not measured by NANOGrav; therefore, its orbital parameters were sourced from \cite{hulse-taylor}, and we assumed a TOA precision of $10 \, \mu\text{s}$. This precision yields a variance for $\delta \dot{P}_b$ according to \cite{BlandfordTeukolsky1976} of the same order of magnitude of the error in the determination of $\dot{P}_b$ reported in the former reference. 

\begin{table*}[t!]
\centering
\small
\setlength{\tabcolsep}{4pt} 
\begin{tabular}{lcccccccccc}
\toprule
Pulsar & $P_b$ [d] & $x$ [lts] & $e$ & $M_p$ [$M_\odot$] & $M_c$ [$M_\odot$] & $\iota$ [$^\circ$] & $\dot{n}$ [yr$^{-1}$] & $\epsilon$ [$\mu$s] & $T_{\mathrm{asc}}$ (MJD) & $T_{\mathrm{obs}}$ [yr] \\
\midrule
J1909$-$3744 & 1.53 & 1.90 & \(1.09\times10^{-7}\) & 1.49 & 0.21 & 86.39 & 2038 & 0.49 & 53292.0 & 15.47 \\
J1012$+$5307 & 0.60 & 0.58 & \(1.28\times10^{-6}\) & 1.35$^{\dagger}$ & 0.12 & 60$^{\dagger}$ & 1664 & 3.39 & 53267.4 & 15.53 \\
J1713$+$0747 & 67.83 & 32.34 & 7.42 $\times 10^{-5}$ & 1.35$^{\dagger}$ & 0.32 & 60$^{\dagger}$ & 1381 & 0.21 & 53420.5 & 15.47 \\
J2145$-$0750 & 6.84 & 10.16 & \(1.93\times10^{-5}\) & 1.35$^{\dagger}$ & 0.50 & 60$^{\dagger}$ & 1202 & 2.36 & 53267.2 & 15.54 \\
J1614$-$2230 & 8.69 & 11.29 & \(1.33\times10^{-6}\) & 1.93 & 0.49 & 89.20 & 1598 & 2.39 & 54724.9 & 11.54 \\
J0613$-$0200 & 1.20 & 1.09 & \(4.19\times10^{-6}\) & 2.97 & 0.24 & 63.92 & 1138 & 1.22 & 53448.8 & 15.04 \\
J2317$+$1439 & 2.46 & 2.31 & \(4.74\times10^{-7}\) & 1.35$^{\dagger}$ & 0.20 & 60$^{\dagger}$ & 890 & 1.64 & 53355.9 & 15.65 \\
J0740$+$6620 & 4.77 & 3.98 & \(5.94\times10^{-6}\) & 2.00 & 0.25 & 87.69 & 2123 & 2.94 & 56640.4 & 6.31 \\
J1738$+$0333 & 0.35 & 0.34 & \(1.12\times10^{-6}\) & 1.35$^{\dagger}$ & 0.10 & 60$^{\dagger}$ & 819 & 3.03 & 55135.8 & 10.73 \\
J1125$+$7819 & 15.36 & 12.19 & \(1.29\times10^{-5}\) & 1.35$^{\dagger}$ & 0.33 & 60$^{\dagger}$ & 1386 & 7.61 & 56640.4 & 6.30 \\
B1855$+$09   & 12.33 & 9.23 & \(2.17\times10^{-5}\) & 1.53 & 0.26 & 87.35 & 498 & 1.23 & 53358.7 & 15.59 \\
J2234$+$0944 & 0.42 & 0.07 & \(6.65\times10^{-6}\) & 1.35$^{\dagger}$ & 0.02 & 60$^{\dagger}$ & 1061 & 2.75 & 56458.4 & 7.10 \\
J2043$+$1711 & 1.48 & 1.62 & \(5.03\times10^{-6}\) & 1.59 & 0.19 & 82.08 & 817 & 1.03 & 55760.3 & 9.05 \\
J1802$-$2124 & 0.70 & 3.72 & \(3.38\times10^{-6}\) & 1.35$^{\dagger}$ & 0.98 & 60$^{\dagger}$ & 1966 & 6.13 & 57682.9 & 3.46 \\
J1719$-$1438 & 0.09 & 0.002 & \(3.96\times10^{-4}\) & 1.35$^{\dagger}$ & 0.001 & 60$^{\dagger}$ & 1846 & 13.18 & 57682.9 & 3.44 \\
J1741$+$1351 & 16.34 & 11.00 & \(9.98\times10^{-6}\) & 0.98 & 0.20 & 74.68 & 507 & 1.33 & 55042.0 & 11.01 \\
J1811$-$2405 & 6.27 & 5.71 & \(1.17\times10^{-6}\) & 2.14 & 0.33 & 73.78 & 1523 & 2.39 & 57682.9 & 3.46 \\
J0610$-$2100 & 0.29 & 0.07 & \(1.54\times10^{-5}\) & 1.35$^{\dagger}$ & 0.02 & 60$^{\dagger}$ & 1451 & 8.38 & 57713.3 & 3.37 \\
J2017$+$0603 & 2.20 & 2.19 & \(6.99\times10^{-6}\) & 2.02 & 0.25 & 67.90 & 422 & 1.77 & 55989.6 & 8.32 \\
J1745$+$1017 & 0.73 & 0.09 & \(1.67\times10^{-5}\) & 1.35$^{\dagger}$ & 0.02 & 60$^{\dagger}$ & 664 & 3.00 & 57390.6 & 4.55 \\
J0406$+$3039 & 6.96 & 2.32 & \(1.24\times10^{-5}\) & 1.35$^{\dagger}$ & 0.10 & 60$^{\dagger}$ & 686 & 1.67 & 57755.1 & 3.56 \\
J0509$+$0856 & 4.91 & 2.46 & \(2.21\times10^{-5}\) & 1.35$^{\dagger}$ & 0.13 & 60$^{\dagger}$ & 609 & 6.04 & 57755.1 & 3.56 \\
J0557$+$1551 & 4.85 & 4.05 & \(8.64\times10^{-6}\) & 1.35$^{\dagger}$ & 0.20$^{\dagger}$ & 60$^{\dagger}$ & 115 & 4.81 & 57389.2 & 4.57 \\
\bottomrule
\end{tabular}
\caption{Observational and orbital parameters corresponding to binary pulsars with low eccentricity. The superscript $^\dagger$ corresponds to assumed values.}
\label{tab_pulsar_data}
\end{table*}
\begin{table*}[t!]
\centering
\small
\setlength{\tabcolsep}{3pt}
\begin{tabular}{lccccccccccc}
\toprule
Pulsar & $P_b$ [d] & $x$ [lts] & $\omega$ [°] & $e$ & $M_1$ [$M_\odot$] & $M_2$ [$M_\odot$] & $\iota$ [°] & $\dot{n}$ [yr$^{-1}$] & $\epsilon$ [$\mu$s] & $T_0$ [MJD] & $T_{\mathrm{obs}}$ [yr] \\
\midrule
J1946$+$3417 & 27.0 & 13.9 & 223.4 & 0.134 & 1.83 & 0.266 & 76.4 & 827.2 & 3.47 & 58008.3 & 5.7 \\
J2234$+$0611 & 32.0 & 13.9 & 277.2 & 0.129 & 1.353 & 0.298 & $138.7$ & 545.5 & 3.25 & 57850.1 & 6.5 \\
J1903$+$0327 & 95.2 & 105.6 & 141.7 & 0.437 & $1.67$ & $1.03$ & $77.35$ & 638.8 & 6.24 & 57109.4 & 10.7 \\
B1913$+$16   & 0.32 & 2.34 & 292.5 & 0.617 & 1.438 & 1.390 & 47.11 & 298.6 & $10.00^{\dagger}$ & 52144.9 & 31.0 \\
\bottomrule
\end{tabular}
\caption{Observational and orbital parameters corresponding to binary pulsars with high eccentricity.}
\label{tab:pulsars_high_eccentricity}
\end{table*}

\twocolumngrid
\section{One-step method}
\label{ap:one-step}

In the main text, we obtained the sensitivity through a two-step method that accounts for the marginalization over the field amplitude, mass and phase as well as other secular effects which may leave an imprint on our measured times-of-arrival. Here, we will compare our results to those obtained through a one-step method, directly fitting the time residuals to the timing model. We assume that the variations in the pulses' times of arrival are exclusively due to the presence of dark matter, and that the orbital parameters are subject to a white gaussian noise distribution.

We will consider the fact that the dominant contribution towards the time residual is given by $\Theta'$, therefore
\begin{equation}
    R^{\text{BT}}_{\Theta'}= - \frac{\delta \Theta' (\alpha_b \sin E' - \eta_b \cos E')}{1 - e \cos E'}.
    \label{eq: time residual BT solo theta'}
\end{equation}

The noise correlation matrix can be written 
\begin{equation}
C_{a,i;b,j} = \epsilon_a^2 \, \delta_{ab}\delta_{ij},
\end{equation}
where $\epsilon_a$ is the noise amplitude associated to pulsar $a$, while $i,j$ go through the TOAs. 

For a given pulsar, the likelihood function of the time residuals, $\{R_i\}_a$, is given by
\begin{equation}
\mathcal{L}_a(\{R_i\}_a|\mathbf{p}_a) = 
\frac{1}{(2\pi)^{n_a/2}\epsilon_a^{n_a}}
\exp\!\left[
-\frac{1}{2}\sum_{i=1}^{n_a}
\frac{(R_{ai}-R^{\mathrm{DM}}_{ai})^2}{\epsilon_a^2}
\right],
\end{equation}
where the model parameters are $\mathbf{p}_a = (\beta, \varrho_a, \Upsilon_a)$, and $R^{\mathrm{DM}}_{ai}$ is the theoretical residual associated with the dark matter signal.

The total combined likelihood of all pulsars is obtained as the product of the individual ones:
\begin{equation}
\mathcal{L}(\{R_{ai}\}|\mathbf{p}) = \prod_{a=1}^{N_p} \mathcal{L}_a(\{R_i\}_a|\mathbf{p}_a).
\end{equation}

Applying Bayes' theorem,
\begin{equation}
\mathcal{L}(\mathbf{p}|\{R_{ai}\}) = 
\frac{\mathcal{L}(\{R_{ai}\}|\mathbf{p}) \, \Pi_0(\mathbf{p})}{P_0(\{R_{ai}\})},
\end{equation}
where $\Pi_0(\mathbf{p})$ represents the prior over the unknown parameters and $P_0(\{R_{ai}\})$ is the evidence. Following the delta-prior approach, we assume that the variables $\varrho_a$ and $\Upsilon_a$ are fixed at their fiducial values, which implies that the only free variable is the coupling $\beta$.

Integrating over the remaining parameters, it is found that the posterior probability for $\beta$ adopts a Gaussian distribution:
\begin{equation}
\mathcal{L}(\beta|\{R_{ai}\}) = 
\frac{1}{\sqrt{2\pi}\,\sigma_\beta}
\exp\!\left[-\frac{(\beta - \beta_{\mathrm{peak}})^2}{2\sigma_\beta^2}\right],
\end{equation}
where the peak and the standard deviation for a single pulsar are given by
\begin{equation}
    \beta_{\mathrm{peak},a} \equiv 
\frac{\mathbf{R}_a \cdot \mathbf{Q}_a}{\mathbf{Q}_a^2} 
\qquad \text{and} \qquad
\sigma_{\beta,a} = \frac{\epsilon_a}{\sqrt{\mathbf{Q}_a^2}}.
\label{eq: fisher - sigma sin marginalizar}
\end{equation}

Here, $\mathbf{Q}_a \equiv \mathbf{R}_a^{\mathrm{DM}}/\beta$ is a vector representing the functional form of the signal induced by ULDM in pulsar $a$ over the observation time, whose amplitude is independent of the value of $\beta$ but depends on the orbital and phase parameters.

Finally, the sensitivity limit is defined by requiring that the detection is distinguishable from the null case with a significance of $3\sigma$,
\begin{equation}
|\beta_f| > 3\,\sigma_\beta,
\end{equation}
which establishes the lower bound on the coupling for which the signal would be statistically detectable.

At this point, it is important to clarify a specific aspect. We have previously stated that in this section, we assume the total time residuals are due to the ULDM effect, neglecting potential secular variations induced by other effects, such as those included in Eq. \eqref{eq: resultado delta a sobre a}. In that context, we incorporated a constant correction to $a$, denoted as $\delta a_0$, and a linear one with time as $\dot{a}_0 \, t$. Since the time integral of $\frac{\delta a}{a}(t)$ appears in Eq. \eqref{eq: cuadratico delta theta}, these terms lead to linear and quadratic contributions, respectively.

These terms, along with the linear terms in Eq. \eqref{eq: cuadratico delta theta} that naturally arise when considering the coupling with ULDM, were absorbed into the nuisance parameters; consequently, in the two-step method, they did not contribute to the sensitivity after marginalization. As described in the main text, 
$\dot{a}_0 $ may also result from well-understood effects that are assumed to be either negligible or independently measurable and removable, such that marginalization over this parameter is unnecessary. The linear terms, however,  involving the error in the determination of ${a}_0$, cannot be independently determined. In the one-step method, these linear terms result in significant contributions over the observation time, leading to more optimistic bounds for the coupling constant.

Therefore, a more general case would involve considering the existence of this parameter $\delta a_0$, which introduces linear variations over time. In this scenario, the likelihood now depends on both parameters: $\beta$ and $\delta a_0$.

We model the time residuals from Eq. \eqref{eq: time residual BT solo theta'} as
\begin{equation}
\mathbf{m}(\delta a_0,\beta) = 
- \frac{(\alpha_b \sin E' - \eta_b \cos E')}{1 - e \cos E'} 
\bigg(\frac{3 \omega_b}{2a}\,\delta a_0\,\mathbf{t} + \beta\,\tilde{\boldsymbol{h}}\bigg),
\end{equation}
where $\mathbf{t}=(t_1,\dots,t_n)$ represents the secular time dependence associated with the linear term $\delta a_0 t$ in the variation of $\Theta'$, and $\tilde{\boldsymbol{h}}=(\tilde{h}_1,\dots,\tilde{h}_n)$ describes the functional form of the perturbation induced by dark matter, independent of the value of $\beta$. The prefactor multiplying both terms arises directly from the BT model for the time residual in Eq. \eqref{eq: time residual BT solo theta'}.

We then define, for each observation $i$,
\begin{equation}
w_i \equiv 
\frac{\alpha_b \sin E'_i - \eta_b \cos E'_i}{1 - e \cos E'_i},
\end{equation}
so that the contribution can be written as
\begin{equation}
m_i(\delta a_0,\beta) = 
- w_i \left( \frac{3 \omega_b}{2a}\,\delta a_0 t_i + \beta\,\tilde{h}_i \right).
\end{equation}

Calling $\mathbf{r}=(R_1,\dots,R_n)$ the vector of measured residuals and $\epsilon$ the Gaussian uncertainty associated with each observation, the chi-squared takes the form
\begin{equation}
\chi^2(\delta a_0,\beta)
= \frac{1}{\epsilon^2}
\sum_{i=1}^n
\big[r_i - m_i(\delta a_0,\beta)\big]^2
\end{equation}

After computing the second partial derivatives of $\chi^2$ with respect to the parameters $\boldsymbol{\theta}=(\delta a_0,\beta)$, the Fisher matrix takes the form
\begin{equation}
F =
\frac{1}{\epsilon^2}
\begin{pmatrix}
\left(\frac{3 \omega_b}{2a}\right)^2 \!\sum w_i^2 t_i^2 &
\left(\frac{3 \omega_b}{2a}\right)\! \sum w_i^2 t_i \tilde{h}_i \\[6pt]
\left(\frac{3 \omega_b}{2a}\right)\! \sum w_i^2 t_i \tilde{h}_i &
\sum w_i^2 \tilde{h}_i^2
\end{pmatrix}.
\end{equation}

When including and marginalizing over $\delta a_0$, the covariance is obtained from the inverse of the Fisher matrix. By defining the quantities $g_i \equiv w_i\, t_i$ and $q_i \equiv w_i\, \tilde{h}_i$, the marginalized variance of $\beta$ results in the following compact notation:
\begin{equation}
    \sigma_\beta^2 = (F^{-1})_{\beta\beta} 
     = \epsilon^2\, \frac{\mathbf{g}\!\cdot\!\mathbf{g}} {(\mathbf{q}\!\cdot\!\mathbf{q})(\mathbf{g}\!\cdot\!\mathbf{g}) - (\mathbf{g}\!\cdot\!\mathbf{q})^2 } , \label{eq: fisher_marginalizado_compact}
\end{equation}
where $\mathbf{g}=(g_1,\dots,g_n)$ and $\mathbf{q}=(q_1,\dots,q_n)$. This expression represents the uncertainty in the estimation of the coupling constant while accounting for the degeneracy introduced by the unknown secular drift associated with $\delta a_0$.

\begin{figure}[t]
    \centering
    \includegraphics[width=0.99\linewidth]{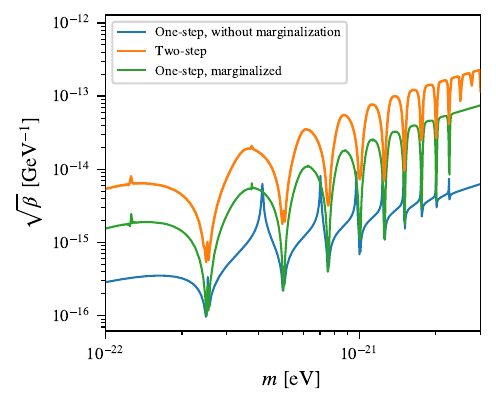}
    \caption{Comparison of the sensitivity obtained through three procedures: the one-step method, the one-step method with marginalization of linear terms, and the two-step method (assuming $\Xi_2=0)$. In each case, the median of ten realizations of the ULDM field parameters is plotted.}
    \label{fig: distintas resonancias}
\end{figure}

Fig. \ref{fig: distintas resonancias} shows the sensitivity curves for the J1903+0327 system obtained using the three methods discussed in this work. First, we can observe that the most stringent and optimistic bound corresponds to the one-step method. This is expected, as this approach assumes that variations in pulse arrival times are due exclusively to the presence of dark matter, without considering other effects that could mimic its signal. For instance, the linear-in-time terms in $\delta \Theta'$ associated with errors in the determination of the semi-major axis are omitted. 

This result is comparable in order of magnitude and general shape to that obtained in \cite{FosterBlas2025}, which also calculated a sensitivity curve showing a significant improvement in the coupling constant bounds near resonant peaks compared to previous estimations such as those in \cite{Blas2017Resonates}. It is worth noting that \cite{FosterBlas2025} assumed a TOA precision of $1\,\mu\text{s}$, while here we employ the uncertainty corresponding to NANOGrav data, $6.24\,\mu\text{s}$, which explains the slightly less stringent bounds.

In the marginalized case, the curves coincide with those of the one-step method at resonant masses but differ away from them. This is because, near resonances, the signal is dominated by oscillatory perturbations, whereas in the non-resonant regime, the secular terms in time that have been marginalized here become relevant.

Finally, the curve obtained through the two-step method lies just above the marginalized one, maintaining a similar general shape. The difference lies in the fact that, in the two-step method, the analysis is performed in two consecutive stages on the timing residuals: first, the standard orbital model parameters are fitted to estimate their associated variances, and then the expected ULDM signal is explicitly incorporated into these already adjusted residuals. This procedure allows for the propagation of orbital parameter uncertainties into the final sensitivity estimate. In contrast, the one-step method performs a single direct fit of arrival times to the complete model, including the potential dark matter contribution to the timing model from the start. For this reason, the bounds obtained via the one-step method are more optimistic, as they do not account for the variance associated with orbital parameter uncertainty.

In summary, the one-step method produces stricter but less realistic limits by not incorporating instrumental uncertainty or secular effects. The marginalized method recovers part of that rigor by accounting for linear variations. Finally, the two-step method, by including both the variance estimation and the full temporal evolution of the perturbations, provides the most physically and statistically consistent description, serving as the most solid basis for combining multiple systems and deriving global bounds on ultralight dark matter coupling.
\section{Sensitivity obtained through the true anomaly}
\label{ap: anomalia}

In the case of eccentric systems, we have studied the sensitivity to $\Theta'$, as was done in \cite{Kus2024}. However, other recent works such as \cite{FosterBlas2025} have directly included the study of the true anomaly $\theta$, introduced in Fig. \ref{fig: orbita kepleriana}. This choice is of particular interest because various sources in the literature prefer using auxiliary variables instead of the true anomaly $\theta$ to avoid a secularly growing term that appears when calculating $\theta(t)$. This approach is even followed in the foundational work of \cite{BlandfordTeukolsky1976}, where $\Theta'$ is the parameter employed for the time residuals in elliptic orbits. We aim to investigate whether using the true anomaly in place of $\Theta'$ leads to discernible variations in the sensitivity curves. To this end, we compare our sensitivity estimates using both variables, applying the one-step method without marginalization described in Appendix \ref{ap:one-step}.

\begin{figure}
    \centering
    \includegraphics[width=0.99\linewidth]{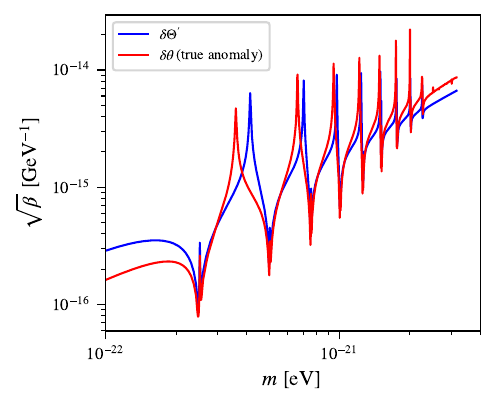}
    \caption{Sensitivity obtained for pulsar J1903+0327 through the true anomaly and the parameter $\Theta'$, for a single realization of the variables $\varrho$ and $\Upsilon$.}
    \label{fig: comparacion anomalia verdadera}
\end{figure}

To this end, we wish to replace $\delta \Theta'$ with $\delta \theta'$ in expression \eqref{eq: time residual BT solo theta'}. From \cite{Gravity}, we have the equation relating the true anomaly $\theta'$ to the eccentric anomaly $E'$:
\begin{equation}
    \tan \bigg( \frac{\theta'}{2}\bigg)= \sqrt{\frac{1 - e}{1+e}} \,  \tan \bigg( \frac{E'}{2}\bigg),
    \label{eq: anomalia y E'}
\end{equation}
as well as the relationship between the mean anomaly $\Theta'$ and the eccentric anomaly $E'$: 
\begin{equation}
    \Theta' = E' - e \sin E'.
    \label{eq: Theta' y E'}
\end{equation}

Before performing the substitution $\delta\Theta' \to \delta\theta'$ in the time residual, two assumptions should be made explicit. First, we assume that $\delta\Theta'$ is the dominant contribution to $R^{\mathrm{BT}}$ across the relevant mass range, so that the full time residual of Eq. \eqref{eq:BT_residual_dominant} is well approximated by Eq. \eqref{eq: time residual BT solo theta'}. This was verified explicitly for the systems considered here, with subdominant terms remaining several orders of magnitude below the $\delta\Theta'$ contribution. Second, we assume that the true anomaly enters the time residual exclusively through $\delta\Theta'$, so that the change of variables $\delta\Theta' \to \delta\theta'$ can be performed directly without modifying other parameters such as $\alpha_b$, $\eta_b$, or $e$.

Under these assumptions, and treating both perturbations to first order, we have
\begin{equation}
    \delta \Theta' = \frac{\partial \Theta'}{\partial \theta'} \, \delta \theta' = \frac{\partial \Theta'}{\partial E'} \, \frac{\partial E'}{\partial \theta'} \, \delta \theta'.
\end{equation}
Differentiating Eqs. \eqref{eq: anomalia y E'} and \eqref{eq: Theta' y E'}, we obtain
\begin{equation}
    \frac{\partial \Theta'}{\partial E'} = 1 - e \cos E'\quad \text{and} \quad \frac{\partial E'}{\partial \theta'}= \frac{\sqrt{1-e^2}}{1+e \cos\theta'},
\end{equation}
so that we finally find
\begin{equation}
    \delta \Theta' = \frac{1 - e \cos E'}{1+ e \cos \theta'} \, \sqrt{1-e^2} \, \delta \theta',
    \label{eq: delta theta' y delta anomalia}
\end{equation}
and by substituting Eq. \eqref{eq: delta theta' y delta anomalia} into Eq. \eqref{eq: time residual BT solo theta'}, we arrive at the expression for the time residual induced by a variation in the true anomaly:
\begin{equation}
    R^{\text{BT}}_{\theta'}= - \frac{\delta \theta' (\alpha_b \sin E' - \eta_b \cos E')}{1 - e \cos E'} \, \frac{1 - e \cos E'}{1+ e \cos \theta'} \, \sqrt{1-e^2} .
    \label{eq: time residual BT solo anomalia}
\end{equation}

To obtain the sensitivity curves, the procedure is identical to the one-step method already described, merely substituting the time residual \eqref{eq: time residual BT solo anomalia} in place of \eqref{eq: time residual BT solo theta'}. For the evolution of the true anomaly \cite{Gravity}, we find 
\begin{equation}
\begin{aligned}
\dot{\theta} &= \left[1+\frac{\beta \Phi^2}{4} - \frac{3}{2} \frac{\delta a}{a} \right] \, \frac{\omega_b}{(1-e^2)^{3/2}}(1 + e \cos \theta)^2 \nonumber \\
&\quad + \frac{1}{e} \left(
    2 \beta \Phi \dot{\Phi} \sin \theta 
    - \omega_b \sqrt{1 - e^2} \, \frac{\beta \Phi^2}{2} \cos \theta
\right),
\end{aligned}
\end{equation}
where numerical integration is preferred over an analytical treatment. The presence of the $(1 + e \cos \theta)^2$ factor implies that a Fourier expansion, such as the one detailed in Appendix \ref{ap: orbital perturb}, would require a prohibitively large number of terms, rendering an analytical solution impractical.

In Fig. \ref{fig: comparacion anomalia verdadera}, we present the sensitivity curves obtained through both variables, for pulsar J1903+0327 with a single realization of $\varrho$ and $\Upsilon$. There, we observe that both curves lie within the same orders of magnitude, and the sensitivity near the resonant peaks is similar in both cases. Therefore, we can conclude that incorporating either variable into the time residual is equivalent if the goal is to estimate the order of magnitude for the coupling constant bound.

\bibliographystyle{apsrev4-2}
\bibliography{referencias_v2}

\end{document}